\documentclass[a4paper]{jpconf}
\usepackage{graphicx}
\usepackage{amssymb,amsmath,cite}

\newcommand{\ba}{\begin{eqnarray}}
\newcommand{\ea}{\end{eqnarray}}
\newcommand{\ban}{\begin{eqnarray*}}
\newcommand{\ean}{\end{eqnarray*}}
\newcommand{\bsub}{\begin{subequations}}
\newcommand{\esub}{\end{subequations}}

\def\ket#1{|#1\rangle}
\def\bra#1{\langle#1|}

\begin{document}
\title{Partial and quasi dynamical symmetries\\ 
in quantum many-body systems}

\author{A. Leviatan}

\address{Racah Institute of Physics, The Hebrew University, 
Jerusalem 91904, Israel}

\ead{ami@phys.huji.ac.il}

\begin{abstract}
We introduce the notions of partial dynamical symmetry (PDS) 
and quasi dynamical symmetry (QDS) and demonstrate their 
relevance to nuclear spectroscopy, to quantum phase transitions and 
to mixed systems with regularity and chaos.
The analysis serves to highlight the potential role of PDS and QDS towards 
understanding the emergent ``simplicity out of complexity" exhibited by 
complex many-body systems.
\end{abstract}

\section{Introduction}

The concept of dynamical symmetry (DS)
is now widely recognized to be of central importance
in our understanding of complex many-body systems.
It had major impact on developments in diverse areas of physics, 
including, hadrons, nuclei and molecules~\cite{BNB,ibm,vibron}.
Its basic paradigm is to write the Hamiltonian of the system under 
consideration in terms of Casimir operators of a chain of nested algebras, 
$G_0\supset G_1 \supset \ldots \supset G_n$. 
The following properties are then observed. 
(i)~All states are solvable and analytic expressions
are available for energies and other observables. 
(ii)~All states are classified by quantum numbers, 
$\vert\alpha_0,\alpha_1,\ldots,\alpha_n\rangle$, 
which are the labels of the irreducible representations (irreps) of the 
algebras in the chain. 
(iii)~The structure of wave functions is completely dictated by symmetry
and is independent of the Hamiltonian's parameters.

The merits of a DS are self-evident.
However, in most applications to
realistic systems, the predictions of an exact DS are rarely fulfilled
and one is compelled to break it.
More often one finds that, in a given system, the assumed symmetry is
not obeyed uniformly, {\it i.e.}, is fulfilled by only some states but
not by others. In describing a transition between different structural
phases, the relevant Hamiltonian, in general, involves competing interactions
with incompatible symmetries. The need to address such situations has led
to the introduction of  partial dynamical symmetry 
(PDS)~\cite{AL92,RamLevVan09,lev11} 
and quasi dynamical symmetry 
(QDS)~\cite{Bahri00,Rowe04,Turner05,Rosensteel05}. 
These intermediate-symmetry notions
and their implications for dynamical systems,
are the subject matter of the present contribution.

\subsection{The interacting boson model}
In order to illustrate  the various notions of symmetries and 
demonstrate their relevance, 
we employ the interacting boson model (IBM)~\cite{ibm}, 
widely used in the description of low-lying quadrupole collective states 
in nuclei in terms of $N$ monopole $(s)$ and
quadrupole $(d)$ bosons representing valence nucleon pairs.
The bilinear combinations 
$\{s^{\dag}s,\,s^{\dag}d_{m},\, d^{\dag}_{m}s,\, 
d^{\dag}_{m}d_{m '}\}$ span a U(6) algebra, which 
serves as the spectrum generating algebra. 
The IBM Hamiltonian is expanded in terms of these generators 
and consists of Hermitian, rotational-scalar interactions 
which conserve the total number of $s$- and $d$- bosons, 
$\hat N = \hat{n}_s + \hat{n}_d = 
s^{\dagger}s + \sum_{m}d^{\dagger}_{m}d_{m}$. 
The three dynamical symmetries of the IBM are 
\bsub
\ba
&&\begin{array}{ccccccc}
{\rm U}(6)&\supset&{\rm U}(5)&\supset&{\rm O}(5)&
\supset&{\rm O}(3)\\
\downarrow&&\downarrow&&\downarrow&&\downarrow\\[0mm]
[N]&&\langle n_d \rangle&&(\tau)&n_\Delta& L
\end{array} ~,
\label{chainu5}
\\[2mm]
&&\begin{array}{ccccccc}
{\rm U}(6)&\supset&{\rm SU}(3)&\supset&{\rm O}(3) &&\\
\downarrow&&\downarrow&&\downarrow\\[0mm]
[N]&&\left (\lambda,\mu\right )& K & L &&
\end{array} ~,
\label{chainsu3}
\\[2mm]
&&\begin{array}{ccccccc}
{\rm U}(6)&\supset&{\rm O}(6)&\supset&{\rm O}(5)&
\supset&{\rm O}(3)\\
\downarrow&&\downarrow&&\downarrow&&\downarrow\\[0mm]
[N]&&\langle \sigma \rangle&&(\tau)&n_\Delta& L
\end{array} ~,
\label{chaino6}
\ea 
\esub
where, below each algebra, its associated labels of irreps 
are given. $n_{\Delta}$ and $K$ are multiplicity labels needed 
in the ${\rm O(5)}\supset {\rm O(3)}$ and 
${\rm SU(3)}\supset {\rm O(3)}$ reductions, 
respectively.
These solvable limits correspond to known benchmarks of 
the geometric description of nuclei~\cite{Bohr75}, 
involving vibrational [U(5)], rotational [SU(3)] 
and $\gamma$-soft [O(6)] types of dynamics.

A geometric visualization of the model is obtained by 
an energy surface
\ba
E_{N}(\beta,\gamma) &=& 
\langle \beta,\gamma; N\vert \hat{H} \vert \beta,\gamma ; N\rangle ~, 
\label{enesurf}
\ea
defined by the expectation value of the Hamiltonian in the coherent 
(intrinsic) state~\cite{gino80,diep80}
\bsub
\ba
\vert\beta,\gamma ; N \rangle &=&
(N!)^{-1/2}(b^{\dagger}_{c})^N\,\vert 0\,\rangle ~,\\[2mm]
b^{\dagger}_{c} &=& (1+\beta^2)^{-1/2}[\beta\cos\gamma 
d^{\dagger}_{0} + \beta\sin{\gamma} 
( d^{\dagger}_{2} + d^{\dagger}_{-2})/\sqrt{2} + s^{\dagger}] ~. 
\ea
\label{condgen}
\esub
Here $(\beta,\gamma)$ are
quadrupole shape parameters whose values, $(\beta_{\rm eq},\gamma_{\rm eq})$, 
at the global minimum of $E_{N}(\beta,\gamma)$ define the equilibrium 
shape for a given Hamiltonian. 
The shape can be spherical $(\beta =0)$ or 
deformed $(\beta >0)$ with $\gamma =0$ (prolate), 
$\gamma =\pi/3$ (oblate), 
$0 < \gamma < \pi/3$ (triaxial), or $\gamma$-independent. 
The equilibrium deformations associated with the 
dynamical symmetry limits are 
$\beta_{\rm eq}=0$ for U(5), $(\beta_{\rm eq} =\sqrt{2},\gamma_{\rm eq}=0)$ 
for SU(3) and $(\beta_{\rm eq}=1,\gamma_{\rm eq}\,{\rm arbitrary})$ for O(6). 

One particularly successful approach within the IBM is the extended 
consistent-Q formalism (ECQF)~\cite{Warner83,Lipas85},
which uses the following Hamiltonian
\ba
\hat H_{\rm ECQF}=
\omega\left[(1-\xi)\,\hat n_d-\frac{\xi}{4N}\,
\hat Q^\chi\cdot\hat Q^\chi\right] ~.
\label{eq:Hamiltonian}
\ea
Here $\hat Q^\chi=d^\dag s+s^\dag\tilde d+\chi\,(d^\dag\tilde d)^{(2)}$
is the quadrupole operator, $\tilde{d}_{m}=(-)^{m}d_{-m}$ 
and the dot implies a scalar product. 
$\xi$ and $\chi$ are the sole structural parameters of the model
since $\omega$ is a scaling factor.
The parameter ranges $0\leq\xi\leq1$ and 
$-\frac{\sqrt{7}}{2}\leq\chi\leq0$ 
interpolate between the U(5), O(6) and SU(3) DS limits, 
which are reached for 
$(\xi,\chi)=(0,\chi)$, $(1,0)$, and $(1,-\frac{\sqrt{7}}{2})$, 
respectively.
It is customary to represent the parameter space by a symmetry 
triangle~\cite{Casten83}, shown in Fig.~1, 
whose vertices correspond to these limits. 
The ECQF has been used extensively
for the description of nuclear properties 
and it was found that the vast majority of nuclei 
are best described by ECQF parameters in the interior of the triangle,
away from any DS limit.
\begin{figure}[t]
\includegraphics[width=14pc]{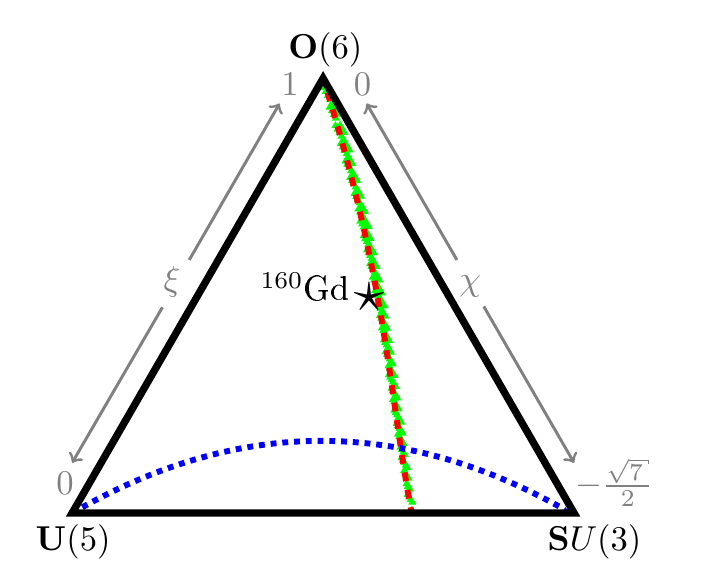}\hspace{2pc}%
\begin{minipage}[b]{20pc}
\caption{
\small
The ECQF symmetry triangle with the position of 
the nucleus $^{160}$Gd indicated by a star. 
The calculated green curve and its approximation by a red dashed line, 
correspond to a region of enhanced purity in the ground state, 
exemplifying an O(6)-PDS discussed in Sections~5.1 and 7.
The blue dotted line shows the ``arc of regularity'' 
mentioned in Section~8, exemplifying SU(3)-QDS.
Adapted from~\cite{kremer14}}.
\label{fig:triangle}
\end{minipage}
\end{figure}
In this context, a key question, addressed in the present contribution, 
can be phrased in the form: are there any remaining ``symmetries'' 
inside the triangle?

\section{Partial dynamical symmetry (PDS)}

In algebraic models, such as the IBM, 
the required symmetry breaking is achieved
by including in the Hamiltonian terms associated with (two or more) 
different sub-algebra chains of the parent spectrum generating algebra. 
In general, under such circumstances, solvability is lost,
there are no remaining non-trivial conserved quantum numbers and all
eigenstates are expected to be mixed.
A partial dynamical symmetry (PDS)~\cite{AL92,RamLevVan09,lev11} 
corresponds to 
a particular symmetry breaking for which some (but not all) of the 
virtues of a dynamical symmetry are retained. 
The essential idea is to relax the stringent conditions
of {\em complete} solvability
so that the properties (i)--(iii) of a DS, mentioned above,
are only partially satisfied.
It is then possible to identify the following types of 
partial dynamical symmetries~\cite{lev11} 
\begin{itemize}
\item {\em PDS type I:} 
$\qquad\;\;\;$ {\bf some} of the states have {\bf all} the 
dynamical symmetry
\item{\em PDS type II:}
$\qquad\;\,$ {\bf all} the states have {\bf part} of the 
dynamical symmetry
\item{\em PDS type III:}
$\qquad$ {\bf some} of the states have {\bf part} of the dynamical 
symmetry.
\end{itemize}
In PDS of type~I, only part of the eigenspectrum is analytically 
solvable and retains all the dynamical symmetry (DS) quantum numbers. 
In PDS of type~II, the 
entire eigenspectrum retains some of the DS quantum numbers. 
PDS of type~III has a hybrid character, in the sense that 
some (solvable) eigenstates keep some of the quantum numbers. 
In what follows we discuss algorithms for constructing Hamiltonians 
with partial dynamical symmetries of various types and demonstrate 
their relevance to quantum many-body systems.

\section{PDS (type I)}

PDS of type I corresponds to a situation for which the defining 
properties of a dynamical symmetry (DS), namely, solvability, 
good quantum numbers, and symmetry-dictated structure are fulfilled exactly, 
but by only a subset of states. 
An algorithm for constructing Hamiltonians with PDS 
has been developed in~\cite{AL92} and further elaborated 
in~\cite{RamLevVan09}. The analysis starts from the chain of nested algebras
\begin{equation}
\begin{array}{ccccccc}
G_{\rm dyn}&\supset&G&\supset&\cdots&\supset&G_{\rm sym}\\
\downarrow&&\downarrow&&&&\downarrow\\[0mm]
[h]&&\langle\Sigma\rangle&&&&\Lambda
\end{array}
\label{chain}
\end{equation}
Eq.~(\ref{chain}) implies that $G_{\rm dyn}$ is the dynamical 
(spectrum generating) algebra of the 
system such that operators of all physical observables 
can be written in terms of its generators; 
a single irrep of $G_{\rm dyn}$
contains all states of relevance in the problem.
In contrast, $G_{\rm sym}$ is the symmetry algebra
and a single of its irreps contains states that are degenerate in energy. 
Assuming, for simplicity,
that particle number is conserved, then 
all states, and hence the representation $[h]$,
can then be assigned a definite particle number~$N$.
For $N$ identical particles the representation 
$[h]$ of the dynamical algebra 
$G_{\rm dyn}$ 
is either symmetric $[N]$ (bosons)
or antisymmetric $[1^N]$ (fermions) and 
will be denoted, in both cases, as $[h_N]$. 
The occurrence of a DS of the type~(\ref{chain})
signifies that its eigenstates can be labeled as
$|[h_N]\langle\Sigma\rangle\dots\Lambda\rangle$;
additional labels (indicated by $\dots$)
are suppressed in the following. 
Likewise, operators can be classified
according to their tensor character under~(\ref{chain})
as $\hat T_{[h_n]\langle\sigma\rangle\lambda}$.

Of specific interest in the construction of a PDS
associated with the reduction~(\ref{chain}),
are the $n$-particle annihilation operators $\hat T$ 
which satisfy the property
\begin{equation}
\hat T_{[h_n]\langle\sigma\rangle\lambda}
|[h_N]\langle\Sigma_0\rangle\Lambda\rangle=0 ~,
\label{anni}
\end{equation}
for all possible values of $\Lambda$
contained in a given irrep~$\langle\Sigma_0\rangle$ of $G$. 
Equivalently, this condition can be phrased in terms of the action 
on a lowest weight (LW) state of the G-irrep $\langle\Sigma_0\rangle$,  
$\hat T_{[h_n]\langle\sigma\rangle\lambda}
|LW;\, [h_N]\langle\Sigma_0\rangle\rangle=0$, 
from which states of good $\Lambda$ can be obtained by projection. 
Any $n$-body, 
number-conserving normal-ordered interaction, 
$\hat{H} = \sum_{\alpha,\beta} 
A_{\alpha\beta}\, \hat{T}^{\dag}_{\alpha}\hat{T}_{\beta}$, 
written in terms of these annihilation operators 
and their Hermitian conjugates (which transform as the
corresponding conjugate irreps), 
can be added to the Hamiltonian with a DS~(\ref{chain}), while
still preserving the solvability of states with 
$\langle\Sigma\rangle=\langle\Sigma_0\rangle$. 
If the operators $\hat T_{[h_n]\langle\sigma\rangle\lambda}$ 
span the entire irrep $\langle\sigma\rangle$ of G, 
then the annihilation condition~(\ref{anni}) is satisfied
for all $\Lambda$-states in $\langle\Sigma_0\rangle$, 
if none of the $G$ irreps $\langle\Sigma\rangle$
contained in the $G_{\rm dyn}$ irrep $[h_{N-n}]$
belongs to the $G$ Kronecker product
$\langle\sigma\rangle\times\langle\Sigma_0\rangle$. 
So the problem of finding interactions
that preserve solvability
for part of the states~(\ref{chain})
is reduced to carrying out a Kronecker product. 
The arguments for choosing the special irrep 
$\langle\Sigma\rangle=\langle\Sigma_0\rangle$ in Eq.~(\ref{anni}), 
which contains the solvable states, are based on 
physical grounds. A~frequently encountered choice is the irrep which 
contains the ground state of the system. 

\subsection{SU(3) PDS (type I) in nuclei}
\label{subsec:su3PDStypeI}

The SU(3) DS chain of the IBM 
and related quantum numbers are given in Eq.~(\ref{chainsu3}).
The DS Hamiltonian involves the Casimir operators 
of SU(3) and O(3), 
with eigenvalues $\lambda^2\!+\!\mu^2\!+\!\lambda\mu\!+\! 
3(\lambda+\mu)$ and $L(L+1)$, 
respectively. 
The spectrum resembles that of an 
axially-deformed rotovibrator and the corresponding  
eigenstates are arranged in SU(3) multiplets. 
The label $K$ corresponds geometrically to the
projection of the angular momentum on the symmetry axis. 
In a given SU(3) irrep $(\lambda,\mu)$, each $K$-value is associated 
with a rotational band and states with the same angular momentum $L$, 
in different $K$-bands, are degenerate. 
The lowest SU(3) irrep is $(2N,0)$, which describes the ground band 
$g(K=0)$ of a prolate deformed nucleus. 
The first excited SU(3) irrep
$(2N-4,2)$ contains degenerate $\beta(K=0)$ and $\gamma(K=2)$ bands. 
This $\beta$-$\gamma$ degeneracy is a characteristic feature of the SU(3) 
limit which, however, is not commonly observed. 
In most deformed nuclei the $\beta$ band lies above the $\gamma$ band. 
In the IBM framework, with at most two-body interactions, one
is therefore compelled to break SU(3) 
in order to conform with the experimental data. 

The construction of Hamiltonians with SU(3)-PDS is based on identification 
of $n$-boson operators which annihilate all states in a given 
SU(3) irrep $(\lambda,\mu)$, 
chosen here to be the ground band irrep $(2N,0)$. 
For that purpose, we consider the following 
two-boson SU(3) tensors, 
$B^{\dagger}_{[n](\lambda,\mu)K;L m}$, with $n=2$, 
$(\lambda,\mu)=(0,2)$ and 
angular momentum $L =0,\,2$
\bsub
\ba
B^{\dagger}_{[2](0,2)0;00} &\propto& 
P^{\dagger}_{0} = d^{\dagger}\cdot d^{\dagger} - 2(s^{\dagger})^2 ~,\\[2mm]
B^{\dagger}_{[2](0,2)0;2m} &\propto& 
P^{\dagger}_{2m} = 2d^{\dagger}_{m}s^{\dagger} + 
\sqrt{7}\, (d^{\dagger}\,d^{\dagger})^{(2)}_{m} ~.
\ea
\label{PL}
\esub 
The corresponding Hermitian conjugate boson-pair annihilation operators,  
$P_0$ and $P_{2m}$, transform 
as $(2,0)$ under SU(3), and 
annihilate all $L$-states in the $(2N,0)$ irrep
\bsub
\ba
P_{0}\,\vert [N](2N,0)K=0, L\rangle &=& 0 ~,
\label{P0cond}\\[2mm]
P_{2m}\,\vert [N](2N,0)K=0, L\rangle &=& 0 ~.
\label{P2cond}
\ea
\label{P0P2}
\esub
Equivalently, these operators 
annihilate 
the coherent state, $\vert\beta=\sqrt{2},\gamma=0 ; N \rangle$, 
of Eq.~(\ref{condgen}), which is the 
lowest-weight state of this irrep 
and serves as an intrinsic state for the SU(3) ground band. 
The relations in Eq.~(\ref{P0P2}) follow from the 
fact that the action of the operators $P_{Lm}$ leads to a state with 
$N-2$ bosons in the U(6) irrep $[N-2]$, 
which does not contain the SU(3) irreps obtained from the product 
$(2,0)\times (2N,0)= (2N+2,0)\oplus (2N,1)\oplus (2N-2,2)$. 
In addition,  $P_0$ satisfies 
\ba
P_{0}\,\vert [N](2N-4k,2k)K=2k, LM \rangle &=& 0 ~, 
\label{P0}
\ea
where for $k> 0$ the indicated $L$-states 
span only part of the SU(3) irreps 
$(\lambda,\mu)=(2N-4k,2k)$ and form the 
rotational members of excited 
$\gamma^{k}(K=2k)$ bands. 

Following the general algorithm, a two-body Hamiltonian with partial 
SU(3) symmetry can now be constructed as~\cite{lev96}
\ba
\hat{H}(h_0,h_2) &=& h_{0}\, P^{\dagger}_{0}P_{0} 
+ h_{2}\,P^{\dagger}_{2}\cdot \tilde{P}_{2} ~,
\label{HPSsu3}
\ea
where $\tilde P_{2m} = (-)^{m}P_{2,-m}$. 
For $h_{2}=h_{0}$, the Hamiltonian is an SU(3) scalar, 
related to the quadratic Casimir operator of SU(3): 
$h_2[-\hat C_{{\rm SU(3)}} + 2\hat N (2\hat N+3)]$.
For $h_0=-5h_2$, it transforms as a $(2,2)$ SU(3) 
tensor component. 
Although in general $\hat{H}(h_0,h_2)$ is not invariant under SU(3), 
Eqs.~(\ref{P0P2})-(\ref{P0}) ensure that it 
retains selected solvable states with good 
SU(3) symmetry, which are 
members of the ground $g(K=0)$ and $\gamma^{k}(K=2k)$ bands with the 
following characteristics
\ba
\begin{array}{lll}
\vert N,(2N,0)K=0,L\rangle & E=0 & L=0,2,4,\ldots, 2N\\[7pt]
\vert N,(2N-4k,2k)K=2k,L\rangle & 
E =  h_{2}\,6k \left (2N - 2k+1 \right ) &
L=K,K+1,\ldots, (2N-2k)~.
\end{array}\;\;
\label{ggamband}
\ea
\begin{figure}[t]
\begin{minipage}{14.1pc}
\includegraphics[height=6cm,width=14.1pc]{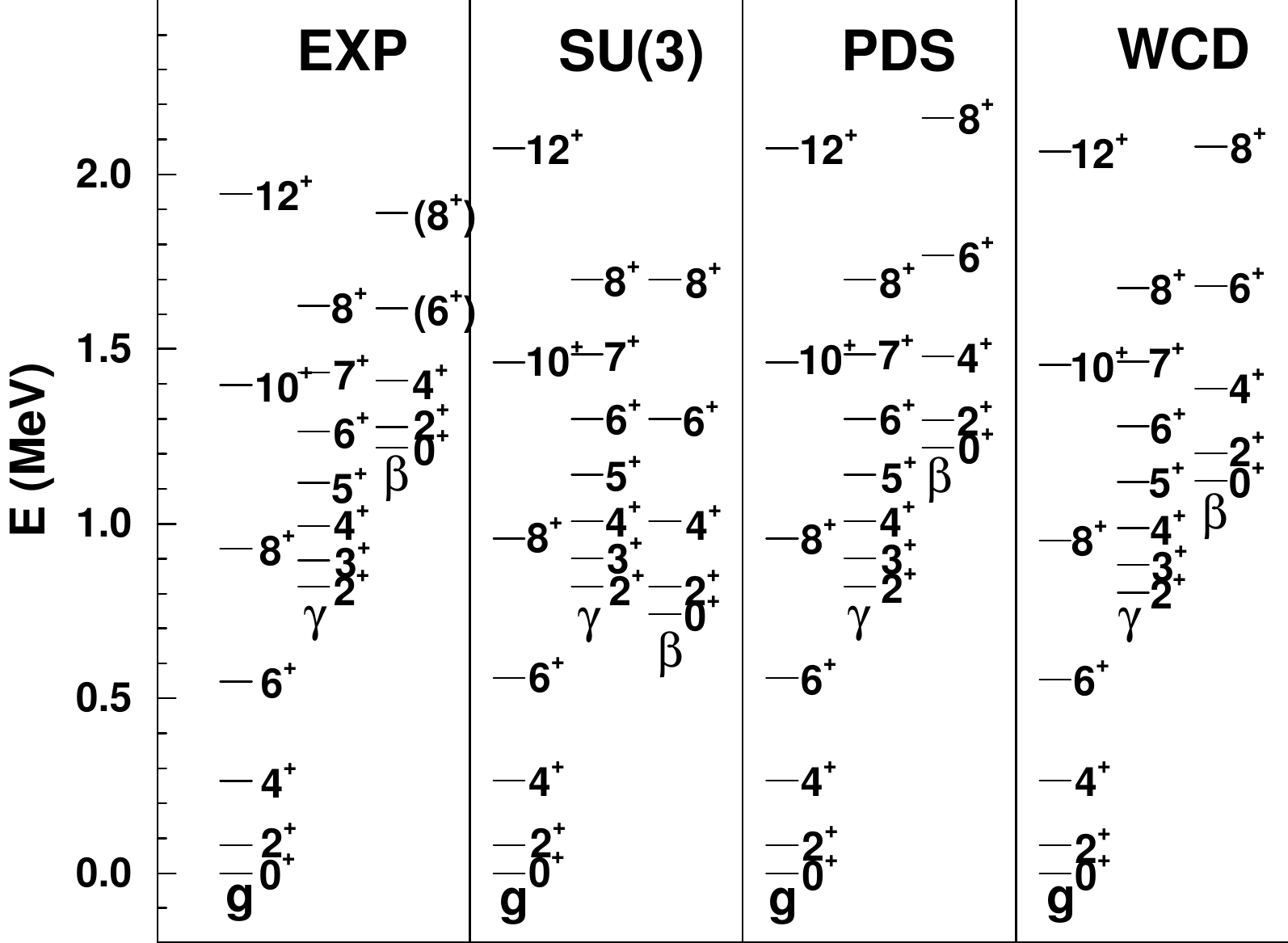}
\caption{
\small
Spectra of $^{168}$Er ($N\!=\!16$). 
Experimental energies
(EXP) are compared with IBM calculations in an exact SU(3) DS [SU(3)], 
in a broken SU(3) symmetry (WCD) 
and in SU(3) PDS. 
The latter employs the Hamiltonian of Eq.~(\ref{HPSsu3}), 
$\hat{H}(h_0,h_2)+C\,\hat{C}_{\rm O(3)}$
with $h_0\!=\!8,\,h_2\!=\!4,\,C\!=\!13$ keV. 
Adapted from ~\cite{lev96}. 
\label{figEr168}}
\vspace{20pt}
\caption{
\small
Comparison of SU(3)-PDS predictions (black bar) with the 
data (red bar) on the relative $\gamma$-band to ground-band 
E2 transitions in several rare earth nuclei. 
Adapted from~\cite{casten14}.}
\end{minipage}\hspace{0.5cm}%
\begin{minipage}{17pc}
\includegraphics[height=14cm]{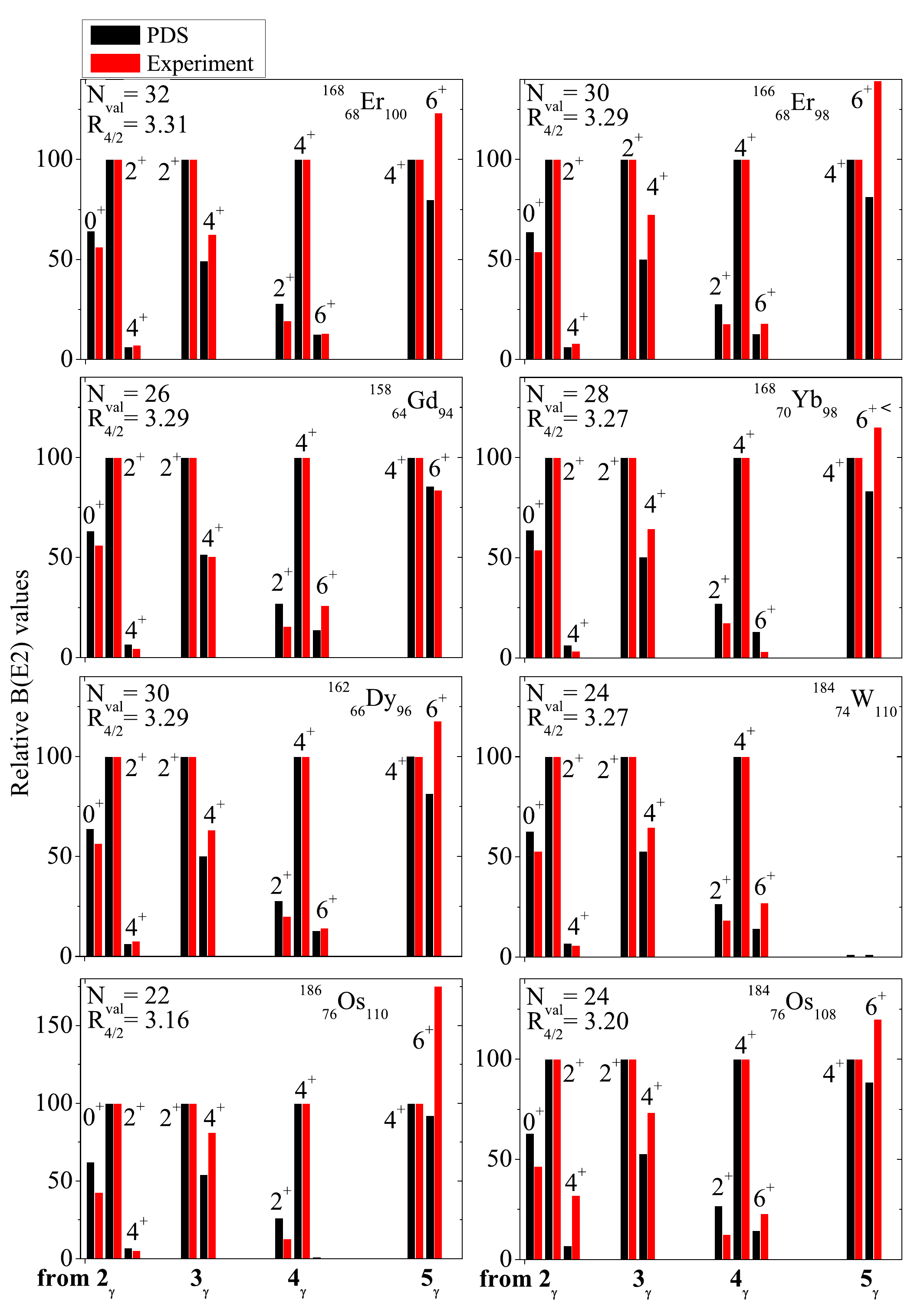}
\end{minipage} 
\end{figure} 
The remaining eigenstates of $\hat{H}(h_0,h_2)$ do 
not preserve the SU(3) symmetry and therefore get mixed. 
One can add to $\hat{H}(h_0,h_2)$ the Casimir operator of O(3), 
$\hat{C}_{{\rm O(3)}}$, which contributes 
an $L(L+1)$ splitting without affecting the wave 
functions. The resulting Hamiltonian has, by construction, SU(3)-PDS.

The empirical spectrum of $^{168}$Er is shown in 
Fig.~\ref{figEr168} and compared with SU(3)-DS, SU(3)-PDS and broken 
SU(3) calculations~\cite{lev96}. The SU(3)-PDS spectrum shows an
improvement over the schematic, exact SU(3) dynamical symmetry 
description, since the $\beta$-$\gamma$ degeneracy is lifted. 
The quality of the calculated PDS spectrum is similar to that obtained
in the broken-SU(3) calculation, however, in the former 
the ground $g(K=0_1)$ and $\gamma(K=2_1)$ bands remain solvable 
with good SU(3) symmetry, $(\lambda,\mu)=(2N,0)$ and $(2N-4,2)$ respectively. 
At the same time, the excited $K=0^{+}_2$ band involves about $13\%$ 
SU(3) admixtures into the dominant $(2N-4,2)$ irrep~\cite{LevSin99}. 
Since the wave functions of the solvable 
states~(\ref{ggamband}) are known, one can obtain 
{\it analytic} expressions for matrix elements of observables between them. 
The SU(3) generator, $\hat{Q}$, is obtained from 
$\hat{Q}^{\chi}$, Eq.~(4), for $\chi=-\frac{\sqrt{7}}{2}$, and hence 
one can write the general E2 operator as $\hat{Q}^{\chi} = \hat{Q} 
+ \theta(d^{\dag}s+s^{\dag}\tilde{d})$. Since $\hat{Q}$
cannot connect different SU(3) irreps, only the second term 
contributes to $\gamma\to g$ transitions. 
Accordingly, the calculated B(E2) ratios for these interband transitions 
are parameter-free 
predictions of SU(3)-PDS. Based on this observation, an extensive test of 
SU(3)-PDS was conducted recently~\cite{casten14}, 
showing evidence for its relevance 
not only for $^{168}$Er but also for a wide range of 
deformed rare earth nuclei. Representative examples of the comparison 
are shown in Fig.~3. Their detailed analysis 
provides insights into the complementary role of finite-nucleon number 
and band-mixing in nuclei.

\section{PDS (type II and type III)}

PDS of type II corresponds to a situation for which {\it all} the
states of the system preserve {\it part} of the dynamical symmetry, 
$G_0 \supset G_1 \supset G_2 \supset \ldots \supset  G_{n}$. 
In this case, there are no analytic solutions,
yet selected quantum numbers (of the conserved symmetries) are retained.
This occurs, for example, 
when the Hamiltonian contains interaction
terms from two different chains with
a common symmetry subalgebra, {\it e.g.}, 
\ba
G_0 \supset 
\left \{
\begin{array}{c}
G_1 \\
G_{1}'
\end{array}
\right \}\supset 
G_2 \supset \ldots \supset G_n ~.
\label{G0chains}
\ea
If $G_{1}$ and $G_{1}'$ are incompatible, {\it i.e.}, do not commute, 
then their irreps are mixed in the eigenstates of the Hamiltonian. 
On the other hand, since $G_2$ and its subalgebras are common to both 
chains, then the labels of their irreps remain as good quantum numbers.  

In the IBM, such a situation arises in Hamiltonians combining terms 
from both the U(5) and O(6) chains, Eqs.~(\ref{chainu5}) 
and~(\ref{chaino6})~\cite{Lev86}. 
All eigenstates now mix U(5) irreps $(n_d)$ and 
O(6) irreps $\langle\sigma\rangle$, but retain the 
$(\tau,L)$ labels of the ${\rm O(5)}\supset {\rm O(3)}$ segment, common 
to both chains, as good quantum numbers. Hamiltonians of this type have 
been used in the study of shape-phase transitions between 
spherical [U(5)] and $\gamma$-soft [O(6)] nuclei~\cite{cjc10}.

An alternative situation where PDS of type II occurs is when the 
Hamiltonian preserves only some of the symmetries $G_i$ in the 
DS chain and only their irreps are unmixed. 
Let $G_1\supset G_2\supset G_3$ be a set of nested algebras which 
may occur anywhere in the chain, 
in-between the spectrum generating algebra $G_0$ and the invariant 
symmetry algebra $G_n$. 
A~systematic procedure~\cite{isa99} 
for identifying interactions with PDS of type II, is based 
on writing the Hamiltonian in terms of generators, $g_i$, of $G_1$, 
which do not belong to its subalgebra $G_2$. 
By construction, such Hamiltonian preserves the 
$G_1$ symmetry but, in general, not the $G_2$ symmetry, and hence will 
have the $G_1$ labels as good quantum numbers but will mix different 
irreps of $G_2$. The Hamiltonians can still conserve the $G_3$ labels 
{\it e.g.}, by choosing it to be a scalar of $G_3$. 
The procedure 
involves the identification of the tensor character under $G_2$ and $G_3$ 
of the operators $g_i$ and their products, $g_i g_{j}\ldots g_{k}$. 
The Hamiltonians obtained in this manner
belong to the integrity basis of $G_3$-scalar operators in 
the enveloping algebra of $G_1$ and, hence, their 
existence is correlated with their order. 

In the IBM, such a scenario can be realized by 
considering an interaction term of the form 
$( (\Pi^{(2)}\times \Pi^{(2)})^{(2)}\cdot\Pi^{(2)}$, 
constructed from the O(6) generator,
$\Pi^{(2)}=d^{\dagger}s+s^{\dagger}\tilde{d}$, 
which is not a generator of O(5)~\cite{isa99}. 
Such a term cannot connect states in different O(6) irreps
but can induce O(5) mixing subject to $\Delta\tau=\pm 1,\pm 3$. 
Consequently, all eigenstates of the resulting Hamiltonian 
have good O(6) quantum number $\sigma$ 
but do not possess O(5) symmetry $\tau$.
These are the necessary ingredients of an O(6) PDS of type II 
associated with the chain of Eq.~(\ref{chaino6}). 

PDS of type III 
has a hybrid character, for which {\it some} of the 
states of the system under study preserve {\it part} 
of the dynamical symmetry~\cite{levisa02}. 
In relation to the dynamical symmetry chain of Eq.~(\ref{chain}), 
with associated basis, $\vert [h_N]\langle\Sigma\rangle\Lambda\rangle$, 
this can be accomplished by relaxing the condition of Eq.~(\ref{anni}), 
so that it holds only for {\it selected} 
states $\Lambda$ contained in a given irrep $\langle\Sigma_0\rangle$ 
of $G$ and/or selected (combinations of) components $\lambda$ of the tensor 
$\hat T_{[h_n]\langle\sigma\rangle\lambda}$. Under such circumstances, 
let $G'\neq G_{sym}$ be a subalgebra of $G$ in the aforementioned 
chain, $G\supset G'$. 
In general, the Hamiltonians, constructed from 
these tensors, in the manner described in Section~3, 
are not invariant under $G$ nor $G'$. 
Nevertheless, they do possess the subset of solvable states, 
$|[h_N]\langle\Sigma_0\rangle\Lambda\rangle$, 
with good $G$-symmetry $\langle\Sigma_0\rangle$ 
(which now span only part of the corresponding $G$-irrep), 
while other states are mixed. At the same time, the symmetry associated 
with the subalgebra $G'$, is broken in all states (including the solvable 
ones). Thus, part of the eigenstates preserve part of the symmetry. 
These are precisely the requirements of PDS of type III.

In the IBM, such a generalized partial symmetry associated with the 
O(6) chain of Eq.~(\ref{chaino6}), can be realized by 
an Hamiltonian constructed of boson-pair operators which are 
not invariant under O(6) nor O(5), but annihilate the 
coherent state, $\vert\beta=1,\gamma=0 ; N \rangle$, 
of Eq.~(\ref{condgen}), which has $\sigma=N$~\cite{levisa02}. 
Such an Hamiltonian has a solvable ground band with good O(6) 
symmetry, which is not preserved by other states. 
All eigenstates, including the solvable ones, break the O(5) symmetry. 
An empirical manifestation of such type of O(6)-PDS is presented
in Section~5.1.

\section{Measures of PDS}

The PDS notion reflects the purity of selected eigenstates with respect 
to a DS basis. The above algorithms 
provide a procedure for an explicit construction 
of Hamiltonians with such property. 
More general (and realistic) Hamiltonians often exhibit features of a 
PDS to a certain approximation. In such cases, one needs to 
assess the quality and applicability of the PDS notion. 
In what follows, we discuss two quantitative measures of PDS, based on 
wave-function entropy and quantum number fluctuations.

Consider an eigenfunction of the IBM Hamiltonian, $\ket{L}$, with 
angular momentum $L$. 
Its expansion in a DS basis reads 
$\ket{L} = \sum_i{C_i\,\ket{[N],\alpha_i,L}}$, 
where $C_i$ stands for the expansion coefficients 
$C^{(L)}_{n_d,\tau,n_{\Delta}}$, $C^{(L)}_{(\lambda,\mu),K}$,
$C^{(L)}_{\sigma,\tau,n_{\Delta}}$, in the U(5), SU(3), O(6) bases, 
Eq.~(1), respectively. 
The probability distributions of U(5): $P_{n_d}^{(L)}$, 
SU(3): $P_{(\lambda,\mu)}^{(L)}$, and O(6): $P_{\sigma}^{(L)}$, 
are calculated as
\bsub
\ba
&&P_{n_d}^{(L)} = \sum_{\tau,n_{\Delta}}
\vert C^{(L)}_{n_d,\tau,n_{\Delta}}\vert^2 \;\;\;\; , \;\;\;\;
S_\mathrm{U5}(L) = 
-\sum_{n_d} P_{n_d}^{(L)} \ln P_{n_d}^{(L)} ~, 
\label{Pnd}
\\[2mm]
&&P_{(\lambda,\mu)}^{(L)} = \sum_{K}
\vert C^{(L)}_{(\lambda,\mu),K}\vert^2 \;\;\;\; , \;\;\;
S_\mathrm{SU3}(L) = 
-\sum_{(\lambda,\mu)} P_{(\lambda,\mu)}^{(L)} \ln P_{(\lambda,\mu)}^{(L)} ~,
\label{Plammu}
\\[2mm] 
&&P_{\sigma}^{(L)} = \sum_{\tau,n_{\Delta}}
\vert C^{(L)}_{\sigma,\tau,n_{\Delta}}\vert^2 \;\;\;\;\;\, , \;\;\;\;
S_\mathrm{O6}(L) = 
-\sum_{\sigma} P_{\sigma}^{(L)} \ln P_{\sigma}^{(L)} ~.
\label{Psigma}
\ea
\label{Shannon}
\esub
The indicated Shannon state entropy $S_{G}(L)$, 
$[G={\rm U(5),\,SU(3),\,O(6)}]$ is a convenient tool to evaluate the purity 
of eigenstates with respect to a DS basis~\cite{CejJol98}.
It vanishes when the considered state 
is pure with good $G$-symmetry
[$S_\mathrm{G}(L)\!=\!0$], and is positive for a mixed state. 
Normalized entropies are obtained by dividing by $\ln D_G$, 
where $D_G$ counts the number of possible $G$-irreps for a given $L$.
In this case, the maximal value [$S_\mathrm{G}(L)\!=\!1$] 
is obtained when $\ket{L}$ is uniformly spread among the irreps of $G$, 
{\it i.e.}, for $P_{G}^{(L)}\!=\!1/D_G$. 
Intermediate values, $0 < S_\mathrm{G}(L) < 1$, 
indicate partial fragmentation 
of the state $\ket{L}$ in the respective DS basis.

The degree of a symmetry of a state $\ket{L}$ can also be inferred 
from the fluctuations of the corresponding quantum number.
As an example, for the O(6) symmetry of the IBM,
the fluctuations in $\sigma$ can be calculated as~\cite{kremer14}
\ba
\Delta\sigma_L=
\sqrt{\sum_i{C_i^2\,\sigma_i^2} 
-\left(\sum_i{C_i^2\,\sigma_i}\right)^2} ~,
\label{eq:fluc2}
\ea
where the sum is over all basis states in the chain, Eq.~(\ref{chaino6}).
If $\ket{L}$ carries an exact O(6) quantum number,
$\sigma$ fluctuations are zero, $\Delta\sigma_L=0$.
If $\ket{L}$ contains basis states with different O(6) quantum numbers,
then $\Delta\sigma_L>0$, indicating that the O(6) symmetry is broken. 
Note that $\Delta\sigma_L$ also vanishes
for a state with a mixture of components with the same $\sigma$
but different O(5) quantum numbers $\tau$,
corresponding to a $\ket{L}$ with good O(6) but mixed O(5) character.
$\Delta\sigma_L$ has the same physical content as 
$S_{O6}(L)$~(\ref{Psigma}) and both can be used as measures of O(6)-PDS.

\subsection{O(6)-PDS (type III) in nuclei}

\begin{figure}[t]
\begin{minipage}{17.5pc}
\includegraphics[width=17.5pc]{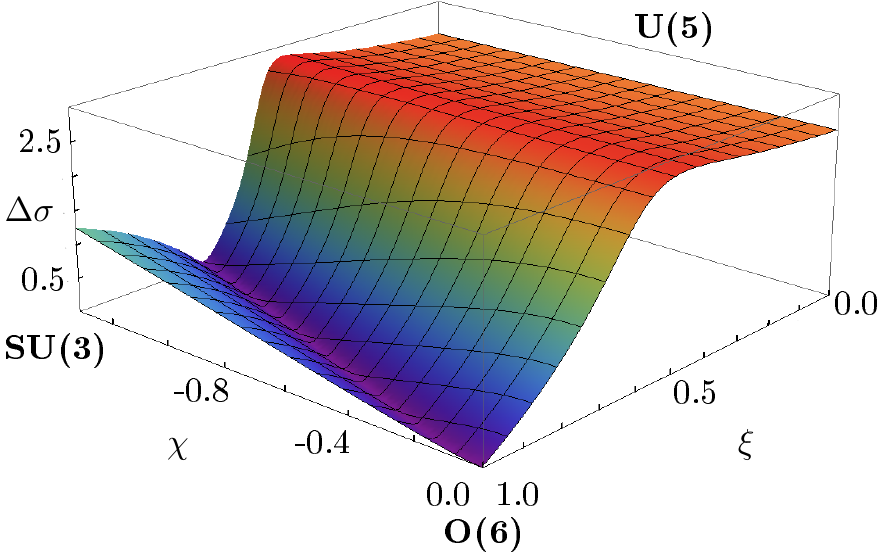}
\caption{\label{fig:3d}
\small
Ground-state ($L\!=\!0^{+}_1$)~fluctuations 
$\Delta\sigma_{0}$~(\ref{eq:fluc2}) 
for $\hat{H}_{\rm ECQF}$~(\ref{eq:Hamiltonian}) 
with $N\!=\!14$. 
The fluctuations vanish at the O(6) DS limit,
saturate towards the U(5) DS limit,
and are of the order $10^{-2}$ in the valley.
Adapted from~\cite{kremer14}.}
\end{minipage}\hspace{2pc}%
\begin{minipage}{17pc}
\includegraphics[width=17pc]{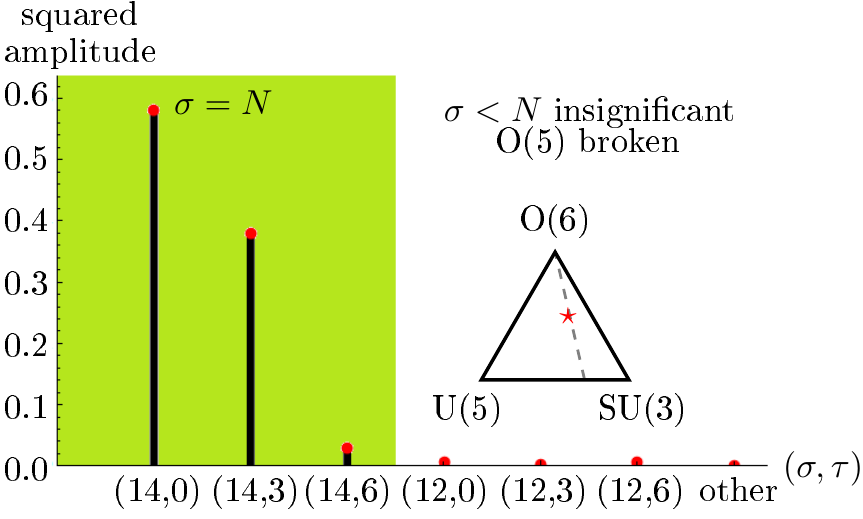}
\caption{\label{fig:wavefunction}
\small
Squared amplitudes $C_i^2$ in the expansion in the O(6) 
basis~(\ref{chaino6}), of the $L\!=\!0^{+}_1$ ground state of 
the ECQF Hamiltonian~(\ref{eq:Hamiltonian}) 
with parameters $\xi\!=\!0.84,\,\chi\!=\!-0.53,\,N\!=\!14$, 
appropriate for $^{160}$Gd. Adapted from~\cite{kremer14}.}
\end{minipage} 
\end{figure}

A recent study~\cite{kremer14} has examined 
the fluctuations $\Delta\sigma_L$, Eq.~(\ref{eq:fluc2}),
for the entire parameter space of the ECQF 
Hamiltonian~(\ref{eq:Hamiltonian}). 
Results of this calculation for the ground state, 
$\ket{L=0^{+}_1}$, with $N=14$, 
are shown in Fig.~\ref{fig:3d}. 
At the O(6) DS limit ($\xi=1$, $\chi=0$), 
$\Delta\sigma_{0}\equiv\Delta\sigma_{L=0_1}$ vanishes per construction
whereas it is greater than zero for all other parameter pairs.
Towards the U(5) DS limit ($\xi=0$),
the fluctuations reach a saturation value of 
$\Delta\sigma_{0}\approx 2.47$. 
At the SU(3) DS limit ($\xi=1$, $\chi=-\frac{\sqrt{7}}{2}$) 
the fluctuations are $\Delta\sigma_{0}\approx 1.25$. 
Surprisingly, one recognizes a valley 
of almost vanishing $\Delta\sigma_{0}$ values,
two orders of magnitude lower than at saturation. 
This region (depicted by a green arc in the triangle of Fig.~1), 
represents a parameter range of the IBM, 
outside the O(6) DS limit, 
where the ground-state wave function
exhibits an exceptionally high degree of purity
with respect to the O(6) quantum number $\sigma$.

The ground-state wave functions in the valley of low $\Delta\sigma_{0}$, 
can be expanded in the O(6)-DS basis (\ref{chaino6}). 
At the O(6) DS limit only one O(6) basis state, with $\sigma=N$ and 
$\tau=0$ contributes, while outside this limit 
the wave function consists of multiple O(6) basis states.
Investigation of the wave function for parameter combinations inside 
the valley reveals an overwhelming dominance of the O(6) basis states 
with $\sigma=N$.
This is seen in Fig.~\ref{fig:wavefunction} for the ground-state 
wave function of $\hat{H}_{\rm ECQF}$~(\ref{eq:Hamiltonian}), 
with parameter values that apply to the nucleus $^{160}$Gd. 
The $\sigma=N$ states comprise 
more than 99\% of the ground-state wave function
at the bottom of the valley
and their dominance causes $\Delta\sigma_{0}$ to be small. 
At the same time, the O(5) symmetry is broken,
as basis states with different quantum number $\tau$
contribute significantly to the wave function. 
Consequently, the valley can be identified
as an entire region in the symmetry triangle
with an approximate O(6)-PDS of type III. 
Outside this valley the ground state is a mixture of several $\sigma$ values
and $\Delta\sigma_{0}$ increases.
\begin{table}[t]
\caption{
\small
Calculated $\sigma$ fluctuations 
$\Delta\sigma_{L}$, Eq.~(\ref{eq:fluc2}), 
for rare earth nuclei in the vicinity of the identified region of 
approximate ground-state-O(6) symmetry~\cite{kremer14}. 
Also shown are the fraction 
$f^{(L)}_{\rm \sigma=N}$ of O(6) basis states with $\sigma = N$ contained 
in the $L\!=\!0,2,4$ states, members of the ground band. 
The structure parameters $\xi$ and $\chi$ are taken 
from \cite{McCutchan04}.\label{nuclei}}
\begin{center}
\lineup
\begin{tabular}{lccccccccc}
\br
Nucleus & $N$ & $\xi$ & $\chi$  
& $\Delta\sigma_{0}$ & $f^{(0)}_{\rm \sigma=N}\;\;$
& $\Delta\sigma_{2}$ & $f^{(2)}_{\rm \sigma=N}\;\;$
& $\Delta\sigma_{4}$ & $f^{(4)}_{\rm \sigma=N}$\cr
\mr
$^{156}$Gd & 12 & 0.72 & -0.86 & 0.46 & 95.3\% & 
0.43 & 95.8\% & 0.38 & 96.6\% \\
$^{158}$Gd & 13 & 0.75 & -0.80 & 0.35 & 97.2\% & 
0.33 & 97.5\% & 0.30 & 97.9\% \\
$^{160}$Gd & 14 & 0.84 & -0.53 & 0.19 & 99.1\% & 
0.19 & 99.2\% & 0.17 & 99.3\% \\
$^{162}$Gd & 15 & 0.98 & -0.53 & 0.41 & 96.0\% & 
0.40 & 96.0\% & 0.40 & 96.1\% \\
$^{160}$Dy & 14 & 0.81 & -0.49 & 0.44 & 96.2\% & 
0.39 & 96.4\% & 0.36 & 96.8\% \\
$^{162}$Dy & 15 & 0.92 & -0.31 & 0.07 & 99.9\% & 
0.07 & 99.9\% & 0.06 & 99.9\% \\
$^{164}$Dy & 16 & 0.98 & -0.26 & 0.13 & 99.6\% & 
0.13 & 99.6\% & 0.13 & 99.6\% \\ 
$^{164}$Er & 14 & 0.84 & -0.37 & 0.39 & 96.5\% & 
0.37 & 96.7\% & 0.35 & 97.1\% \\
$^{166}$Er & 15 & 0.91 & -0.31 & 0.12 & 99.7\% & 
0.11 & 99.7\% & 0.10 & 99.7\% \\
\br
\end{tabular}
\vspace{-12pt}
\end{center}
\end{table}

Detailed ECQF fits for energies and electromagnetic 
transitions of rare-earth nuclei, 
performed in~\cite{McCutchan04}, 
allow one to relate the structure of collective nuclei
to the parameter space of the ECQF Hamiltonian~(\ref{eq:Hamiltonian}). 
From the extracted ($\xi,\chi$) parameters one can calculate
the fluctuations $\Delta\sigma_L$
and the fractions $f_{\sigma=N}$ of squared $\sigma=N$ amplitude. 
Nuclei with $\Delta\sigma_{0} < 0.5$ and $f_{\sigma=N}>95\%$ in the 
ground-state ($L=0^{+}_1$) are listed in Table~\ref{nuclei}.
These quantities are also calculated for yrast states 
with $L>0$ and exhibit similar values in each nucleus. 
It is evident that a large set of rotational rare earth nuclei, 
such as $^{160}$Gd, 
are located in the valley of small $\sigma$ fluctuations. 
They can be identified as candidate nuclei
with an approximate O(6)-PDS of type III 
not only for the ground state, 
but also for the members of the band built on top of it.
\begin{figure}[t]
\begin{minipage}{17pc}
\includegraphics[height=.3\textheight,width=17pc]{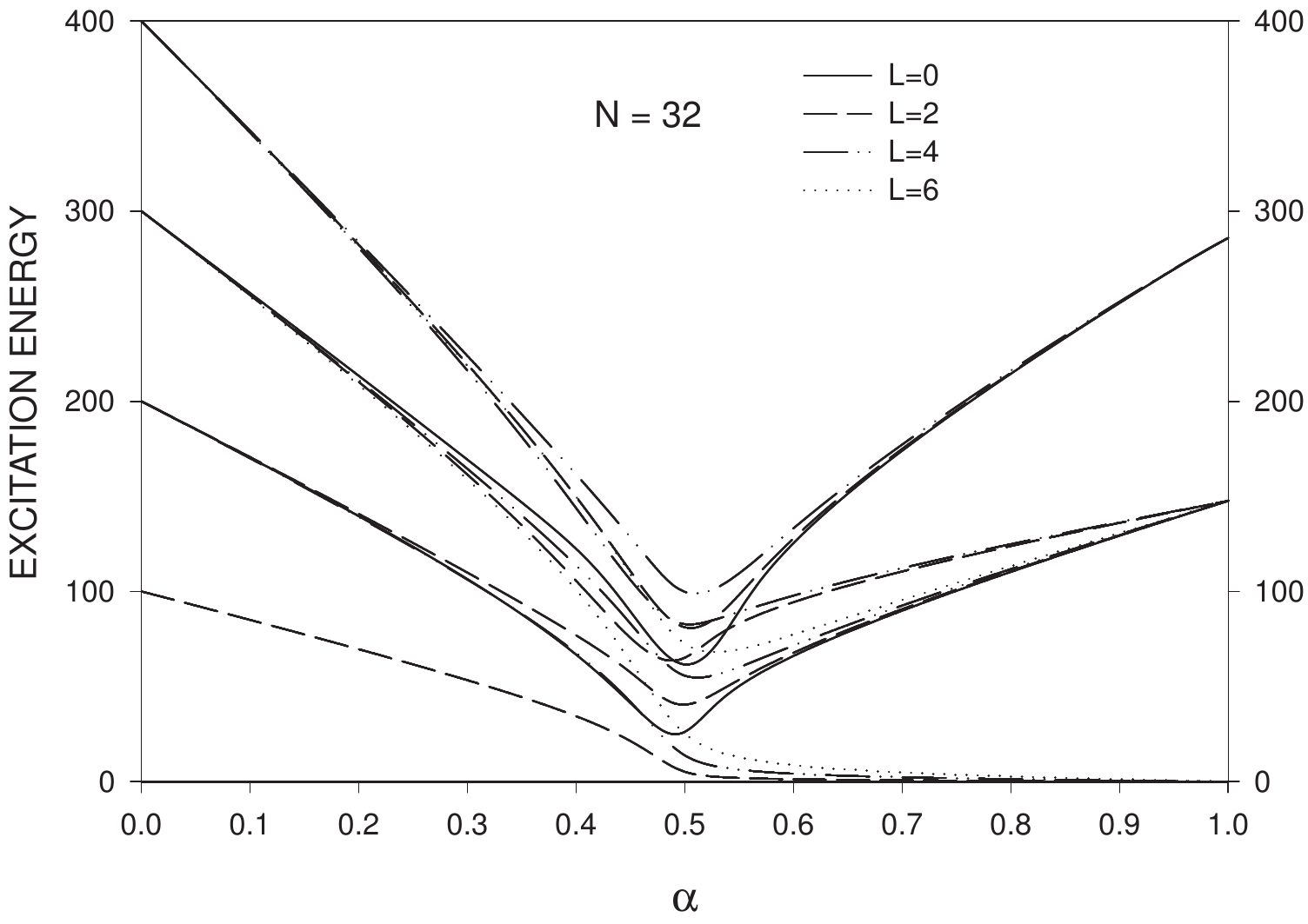}
\caption{\label{fig3a-rowe}
\small
Spectrum of $\hat{H}(\alpha)$, 
Eq.~(\ref{Halpha}), 
interpolating between the U(5) ($\alpha=0$)
and SU(3) ($\alpha=1$) DS limits. From~\cite{Rowe-pc}.} 
\end{minipage}\hspace{2pc}%
\begin{minipage}{18pc}
\vspace{-3cm}
\hspace{-2.2pc}
\includegraphics[height=14cm,width=23pc]{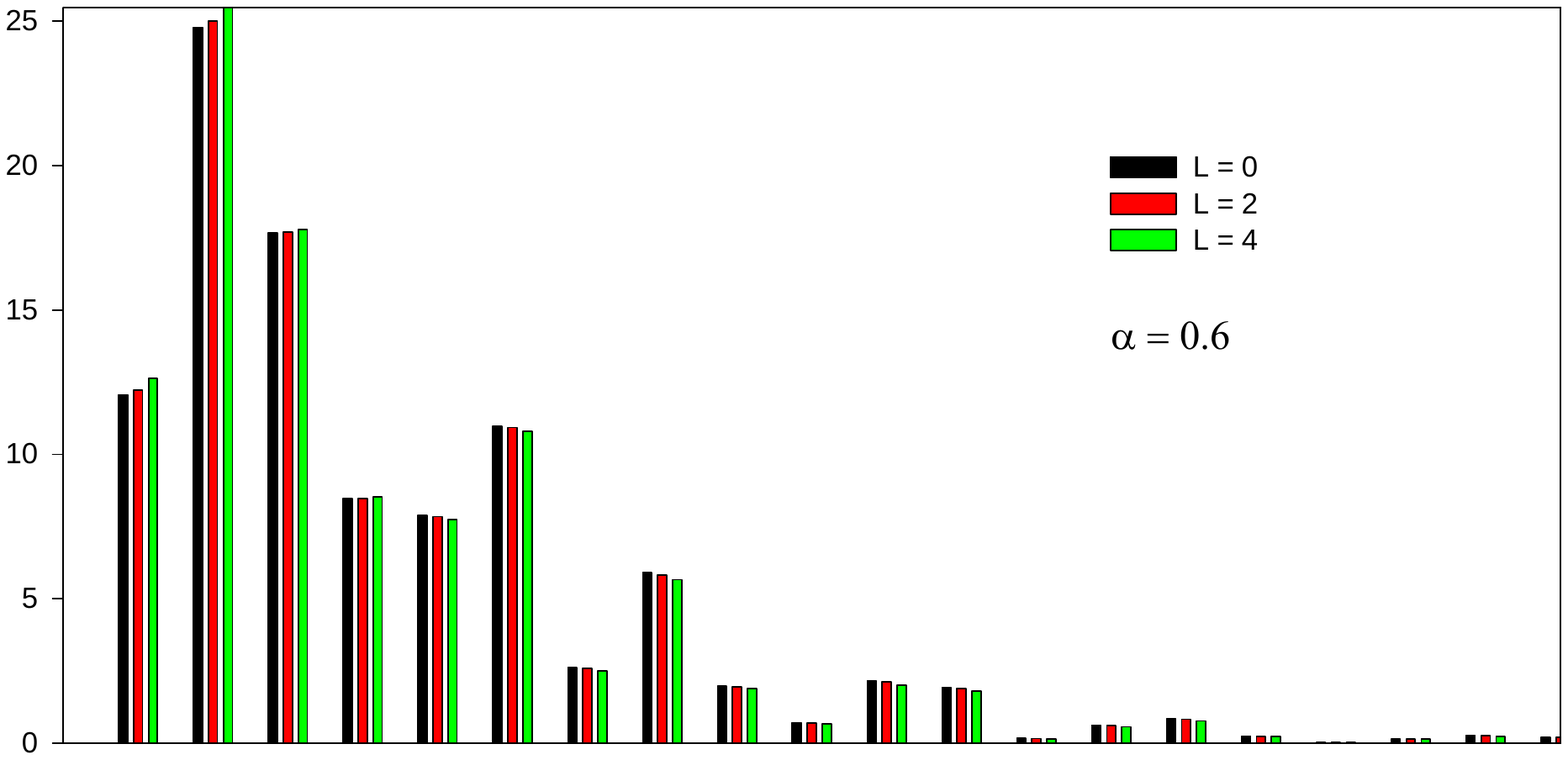}
\vspace{-5cm}
\caption{\label{fig3b-rowe}
\small
Squared amplitudes for angular momenta $L=0,2,4$ yrast 
eigenstates of the U(5)-SU(3) Hamiltonian $\hat{H}(\alpha)$, 
Eq.~(\ref{Halpha}), with $\alpha\!=\!0.6$,
in the SU(3) basis~(\ref{chainsu3}). 
From~\cite{Rowe-pc}}
\end{minipage} 
\end{figure}

\section{Quasi dynamical symmetry (QDS)}

A second kind of intermediate-symmetry occurring in algebraic modeling of 
dynamical systems, is that of quasi dynamical symmetry 
(QDS)~\cite{Bahri00,Rowe04,Turner05,Rosensteel05}. 
While QDS can be defined mathematically in terms of embedded 
representations~\cite{Rowe88,Rowe04b},
its physical meaning is that 
several observables associated with particular eigenstates, 
may be consistent with a certain symmetry which in fact is broken 
in the Hamiltonian. This typically occurs for a Hamiltonian
transitional between two DS limits 
\ba
\hat{H}(\alpha) &=& 
(1-\alpha)\,\hat{H}_{\rm G_1} + \alpha\,\hat{H}_{\rm G_2} ~.
\label{Halpha}
\ea
$\hat{H}(\alpha)$ involves competing incompatible (non-commuting) symmetries. 
For $\alpha=0$ or $\alpha=1$, one recovers the limiting symmetries. 
For $0<\alpha<1$, both symmetries are broken and mixing occurs. 
A detailed study~\cite{Rowe04,Turner05,Rosensteel05}
of such Hamiltonians has found that for most values of $\alpha$, 
selected states continue to exhibit characteristic properties 
({\it e.g.}, 
energy and B(E2) ratios) of the {\it closest} DS limit. 
Such an ``apparent'' persistence of symmetry in the face of strong 
symmetry-breaking interactions, defines a QDS.
The indicated 
persistence is clearly evident in the spectrum shown in Fig.~6, 
for an IBM Hamiltonian, $\hat{H}(\alpha)$, interpolating between 
the $G_1=U(5)$ and $G_2=SU(3)$ DS limits, relevant to shape-phase 
transitions between spherical and axially-deformed nuclei~\cite{cjc10}.
The ``apparent'' symmetry is due to the 
coherent nature of the mixing.
As seen in Fig.~7, the mixing of SU(3) irreps is large, 
but is approximately independent of the angular momentum of 
the yrast states, {\it i.e.}, 
the SU(3) expansion coefficients 
$C^{(L)}_{(\lambda,\mu),K}\approx$ independent 
of $L$. The set of states exhibiting SU(3)-QDS 
thus have a common structure and form a band (the ground band) 
associated with a single intrinsic state. 

The criterion for the validity of QDS is the similarity 
of the decomposition in the given DS basis.
Thus, in the IBM, a quantitative measure of SU(3)-QDS 
can be defined as $\sqrt{1-\bar{\Theta}}$~\cite{NPN14}, 
where $\bar{\Theta}$ is the average of 
$\Theta_{LL'} = \sum_{i}{C_i^LC_i^{L'}}$
for all pairs $L\neq L'$, and $C_i^L$ are the amplitudes of the 
chosen set of eigenstates with angular momentum $L$ in the 
SU(3) basis~(\ref{chainsu3}).

The coherent decomposition signaling SU(3) QDS, implies 
strong correlations between the SU(3) components of different $L$-states 
in the same band. This can be used as an alternative criterion for the 
identification of rotational bands with SU(3)-QDS~\cite{Macek10}. 
Focusing on the $L\!=\!0,2,4,6$, members of $K\!=\!0$ bands,
given a $L=0^{+}_i$ state, among the ensemble of possible states, 
we associate with it those $L_j>0$ states which show the maximum 
correlation, $\max_{j}\{\pi(0_i,L_j)\}$.
Here $\pi(0_i,L_j)$ is a Pearson coefficient whose values lie
in the range $[-1,1]$.  Specifically, 
$\pi(0_i,L_j)=1,-1,0,$ indicate a perfect correlation, 
a perfect anti-correlation, and no linear correlation, respectively, 
among the SU(3) components of the $0_i$ and $L_j$ states. 
To quantify the amount of coherence (hence of SU(3)-QDS) in the chosen 
set of states, one can adapt the procedure of~\cite{Macek10} 
and employ the following product of the maximum 
correlation coefficients:
$C_{\rm SU3}(0_i{\rm -}6) \equiv 
\max_{j}\{\pi(0_i,2_j)\}\,
\max_{k}\{\pi(0_i,4_k)\}\,
\max_{\ell}\{\pi(0_i,6_{\ell})\}$. 
The set of states $\{0_i,\,2_j,\,4_k,\,6_{\ell}\}$ is considered as 
comprising a $K=0$ band with SU(3)-QDS, 
if $C_{\rm SU3}(0_i{\rm -}6)\approx 1$. 
\begin{figure}[t]
\centering
\includegraphics[width=12pc]{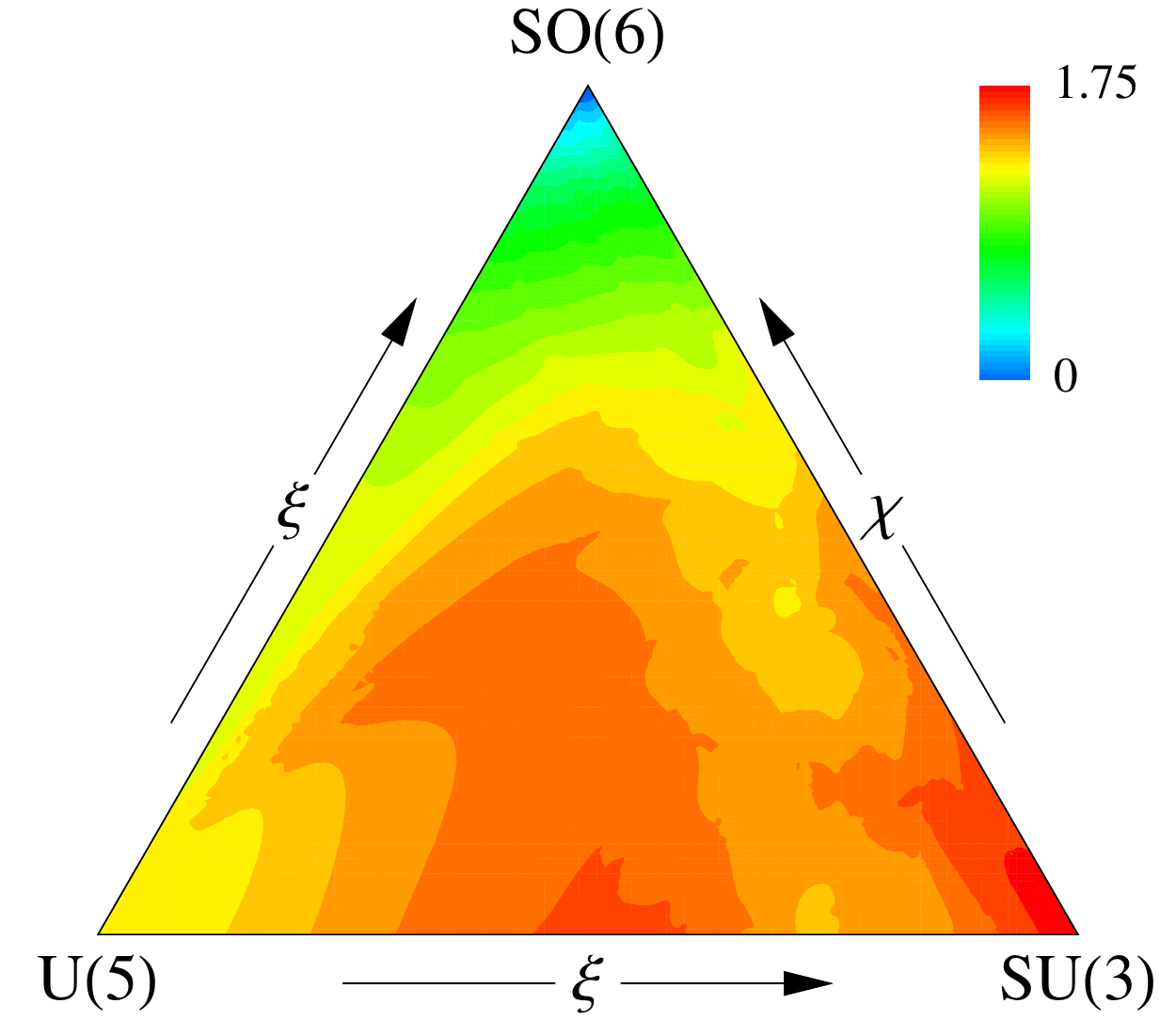}
\includegraphics[width=12pc]{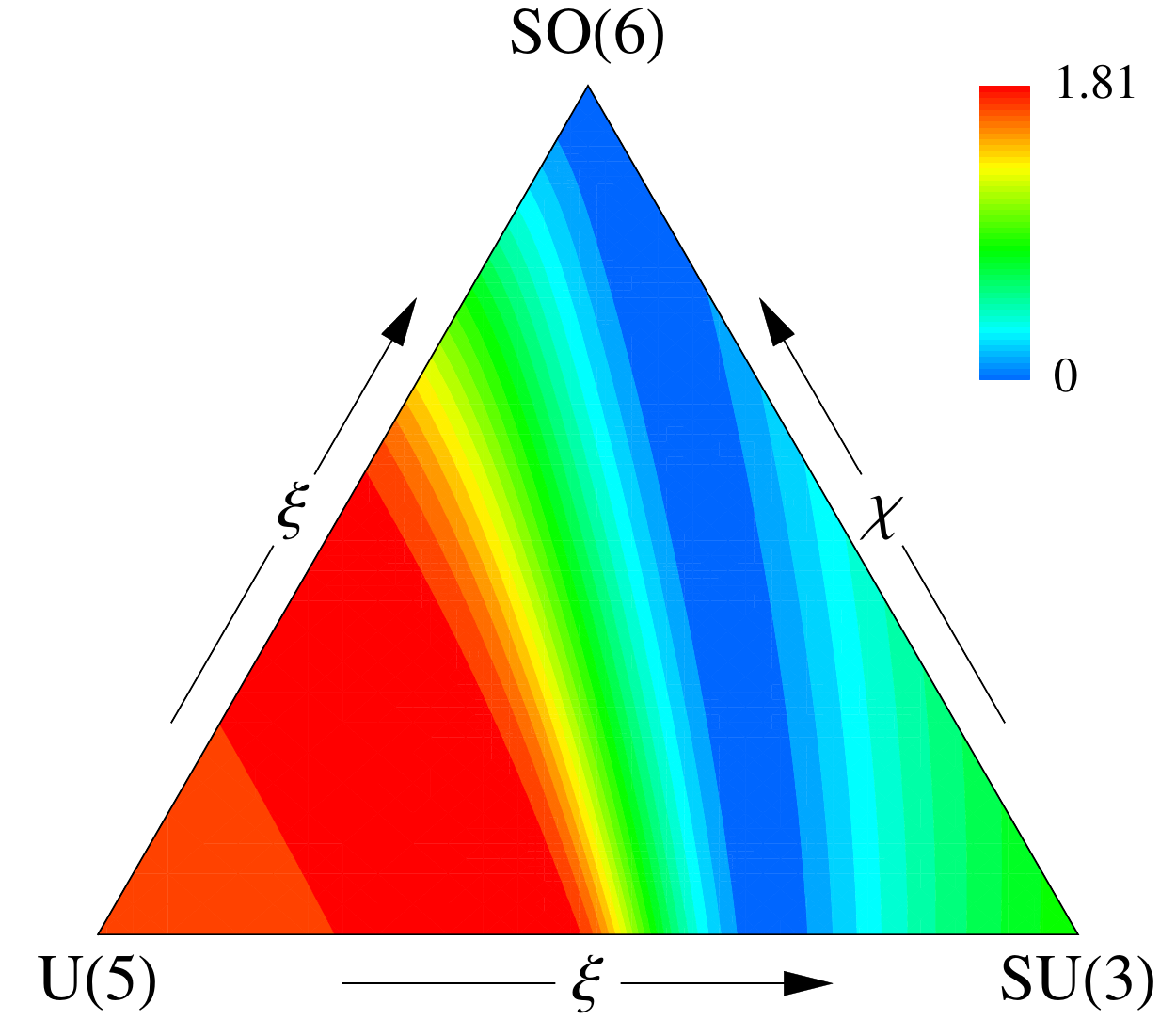}
\includegraphics[width=12pc]{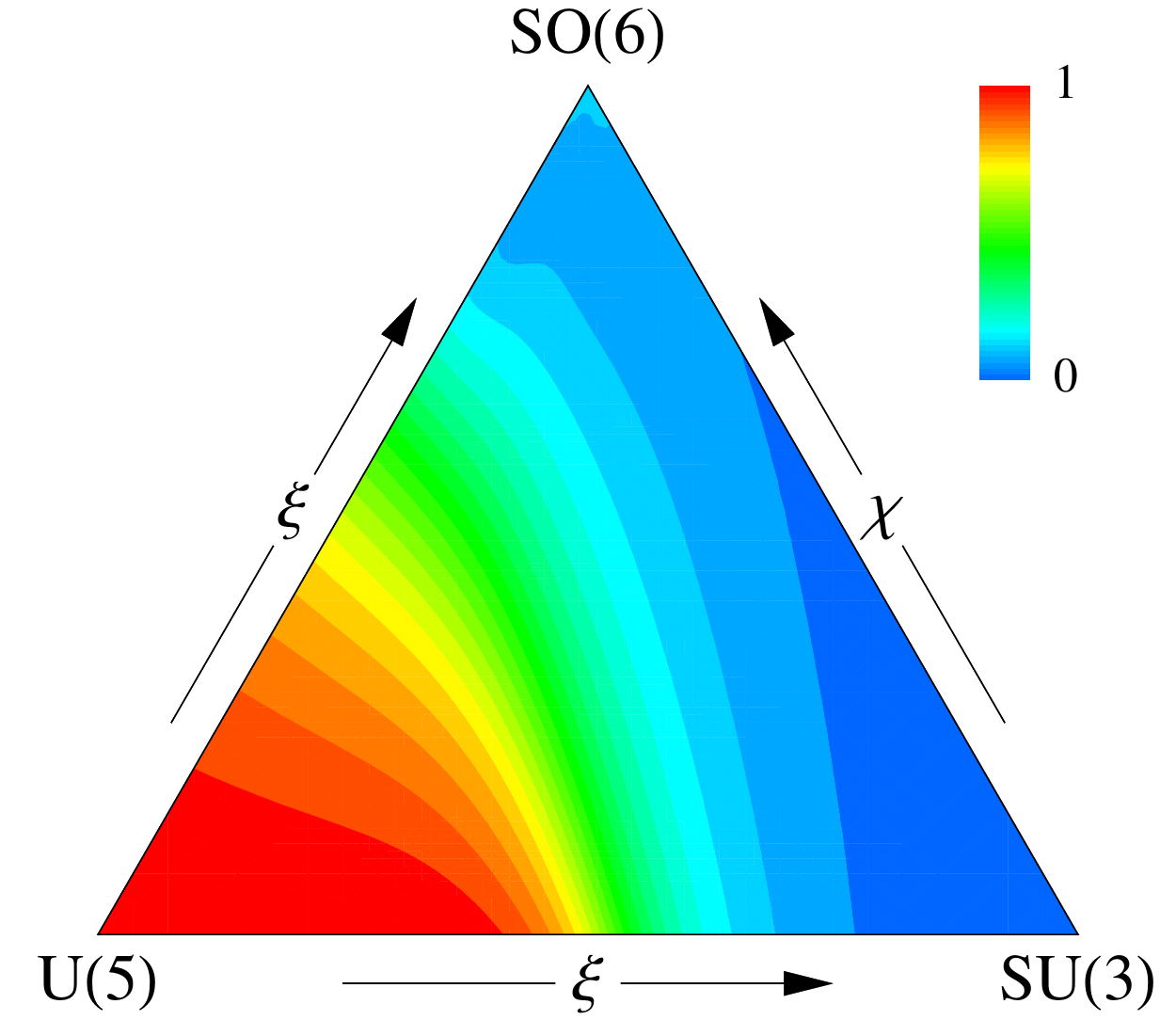}
\caption{\label{fig:pds-qds}
\small
PDS and QDS for 
eigenstates of the ECQF Hamiltonian, 
Eq.~(\ref{eq:Hamiltonian}), with $N\!=\!15$. 
Left~panel: the O(6)-PDS measure [Shannon entropy 
$S_{\rm O6}(L)$, Eq.~(\ref{Psigma})], averaged 
for all $L=0$ eigenstates. Middle~panel: 
$S_{\rm O6}(L)$ for the ground state $L=0_1$. 
Right panel: the quantity $\sqrt{1-\bar{\Theta}}$, 
which is the SU(3)-QDS measure (discussed in the text) 
for yrast (lowest energy) states with $L=0,2,\ldots, 10$. 
From~\cite{Isacker-pc}.}
\end{figure}

\subsection{PDS and QDS in the symmetry triangle}

Recently, a comprehensive analysis of the PDS 
and QDS properties of 
$\hat{H}_{\rm ECQF}$, Eq.~(\ref{eq:Hamiltonian}), 
was carried out~\cite{NPN14}, 
employing the symmetry measures discussed above.
Representative results are shown in Fig.~8. 
The left panel of Fig.~8, shows the O(6) wave function entropy, 
$S_{\rm O6}(L\!=\!0)$, Eq.~(\ref{Psigma}), averaged over 
{\em all} $L=0$ eigenstates. 
As seen, only a small region (marked in blue) 
near the O(6) vertex where an exact O(6) DS occurs, shows a high degree 
of purity with respect to the O(6) quantum number, $\sigma$.
The middle panel displays $S_{\rm O6}(L\!=\!0_1)$ 
of only the lowest $L=0_1$ state. Here, $\sigma$ is 
conserved in the ground state throughout an entire region (marked in blue) 
of ECQF Hamiltonians, reflecting the O(6)-PDS of type~III 
discussed in Section~5.1. The right panel displays the QDS measure
$\sqrt{1-\bar{\Theta}}$, with respect to the SU(3) basis.
It is seen that large areas of the triangle are blue, {\it i.e.},
display SU(3)-QDS. These results illustrate the wider applicability in nuclei 
of the extended concepts, PDS and QDS, as compared to an exact DS.

\section{Linking PDS and QDS}

The concept of PDS reflects the purity of selected states, hence is 
different from the concept of QDS which reflects a coherent mixing. 
Nevertheless, a link between these two hitherto 
unrelated symmetry concepts can be established and shown to be 
empirically manifested in rotational nuclei~\cite{kremer14}.

\begin{figure}[t]
\centering
\includegraphics[width=16cm]{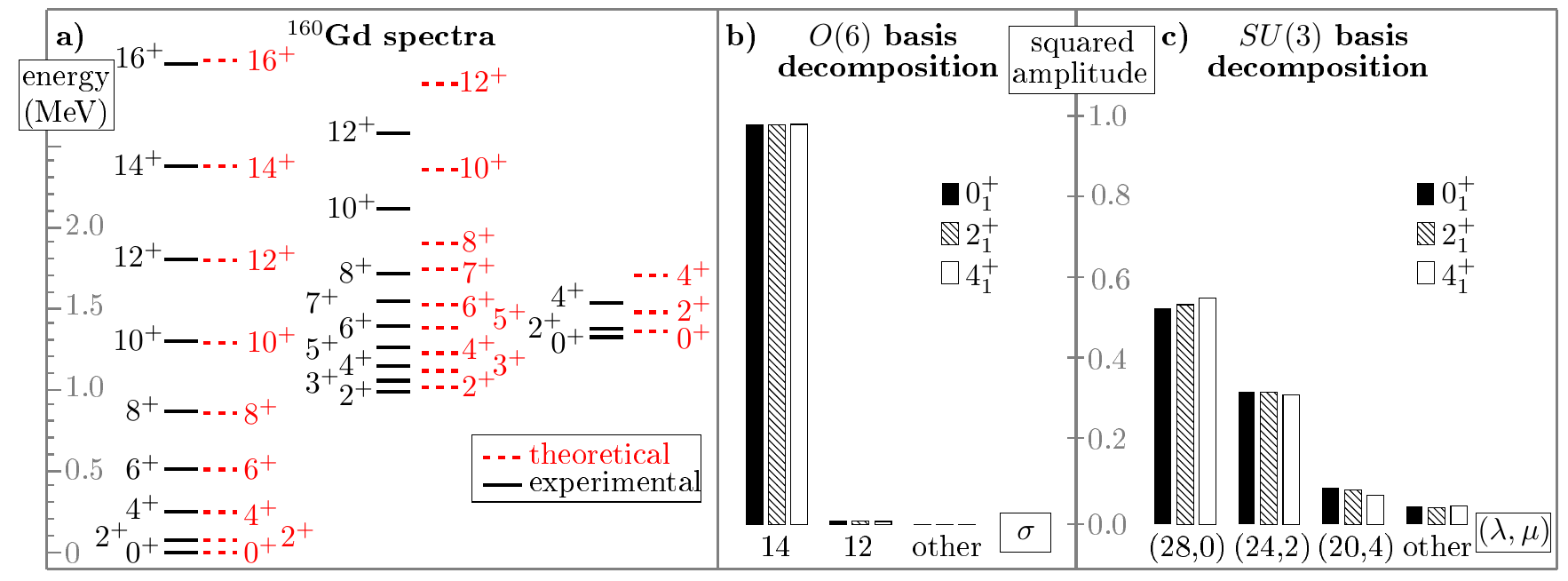}
\caption{
\label{fig:160Gd}
\small
a)~The experimental spectrum of $^{160}$Gd
compared with IBM calculation 
using $\hat{H}_{\rm ECQF}$~(\ref{eq:Hamiltonian}) with 
$\xi\!=\!0.84,\,\chi\!=\!-0.5,\,N\!=\!14$. 
b)~The O(6) decomposition in $\sigma$ components of yrast states 
with $L=0,2,4$.
c)~The SU(3) decomposition in $(\lambda,\mu)$ components of 
the same states. Adapted from~\cite{kremer14}.}
\end{figure}

The experimental spectrum of $^{160}$Gd,
along with its ECQF description~(\ref{eq:Hamiltonian}), 
is shown in the left panel of Fig.~\ref{fig:160Gd}. 
The middle and right panels
show the decomposition into O(6) and SU(3) basis states, respectively,
for yrast states with $L=0,2,4$. 
It is evident that the SU(3) symmetry is broken,
as significant contributions of basis states
with different SU(3) quantum numbers $(\lambda,\mu)$ occur.
It is also clear from Fig.~\ref{fig:160Gd}c
that this mixing occurs in a coherent manner
with similar patterns for the different members of the ground-state band.
As explained in Section~6, this is the hallmark of SU(3) QDS. 
On the other hand, as seen in Fig.~\ref{fig:160Gd}b, 
the yrast states with $L=0,2,4$
are almost entirely composed out of O(6) basis states with $\sigma=N=14$
which implies small fluctuations $\Delta\sigma_L$ (\ref{eq:fluc2})
and the preservation of O(6) symmetry in the ground-state band.
At the same time, as shown in Fig.~\ref{fig:wavefunction}, 
the O(5) symmetry is broken in these states. 
Thus an empirically-manifested link is established between SU(3) QDS and 
O(6) PDS of type~III.

The SU(3) QDS property for the members of the ground band
results from the existence
of a single intrinsic state, which is the coherent state, 
$\vert\beta=1,\gamma=0 ; N \rangle$
of Eq.~(\ref{condgen}) that has $\sigma=N$. 
Indeed, the ($\xi,\chi)$ parameter range of the 
ECQF Hamiltonian for which the equilibrium deformations are 
($\beta=1,\gamma=0$), shown by a red dashed line in Fig.~\ref{fig:triangle}, 
coincides with the region of an approximate 
ground-state O(6) symmetry for large $N$. These results demonstrate that 
coherent mixing of one symmetry (QDS)
can result in the purity of a quantum number
associated with partial conservation of a different, incompatible symmetry 
(PDS).

\section{Impact of PDS and QDS on mixed regular and chaotic dynamics}

Hamiltonians with a dynamical symmetry are always completely integrable. 
The Casimir invariants of the algebras in the chain provide a set 
of constants of the motion in involution. The classical motion is purely 
regular. A symmetry-breaking is connected 
to non-integrability and may give rise to chaotic motion. 
Hamiltonians with PDS and QDS are not completely integrable, 
hence can exhibit stochastic behavior, nor are they completely chaotic, 
since some eigenstates preserve the symmetry exactly in a PDS 
or mix in a coherent fashion in a QDS.
Consequently, Hamiltonians with such intermediate symmetries 
are optimally suitable to the study
of mixed systems with coexisting regularity and chaos. 

The dynamics of a generic classical Hamiltonian system is mixed;
KAM islands of regular motion and chaotic regions coexist
in phase space. 
In the associated quantum system, 
the statistical properties of the spectrum 
are usually intermediate between the Poisson and the Gaussian orthogonal 
ensemble (GOE) statistics. 
In a PDS, the symmetry of the subset of solvable states 
is exact, yet does not arise from invariance properties of the 
Hamiltonian. Several works have shown that a PDS is strongly 
correlated with suppression 
of chaos~\cite{walev93,levwhe96}. 
This enhancement of regularity was seen in both 
the classical measures of chaos, {\it e.g.}, 
the fraction of chaotic volume and the 
average largest Lyapunov exponent, and in quantum measures of chaos, 
{\it e.g.}, the nearest neighbors level spacing distribution, whose 
parameter interpolates between the Poisson and GOE statistics. 
The reduction in chaos occurs even when 
the fraction of solvable states approaches zero in the classical limit, 
suggesting that the existence of a PDS increases the purity of 
other neighbouring states in the system.

The coherent mixing common to a set of states, characterizing a QDS, 
results from the existence of a single intrinsic state for each such 
band and imprints an adiabatic motion and increased regularity~\cite{MDSC10}.
This was verified for low-~\cite{Rosensteel05} and high-lying~\cite{Macek10} 
rotational bands using the ECQF Hamiltonian, Eq.~(\ref{eq:Hamiltonian}).
SU(3) QDS has been proposed~\cite{Bona10} to underly
the ``arc of regularity''~\cite{Alhassid91},
a narrow zone of enhanced regularity in the
parameter-space of $\hat{H}_{\rm ECQF}$. 
The arc, shown by a blue dotted line in Fig.~\ref{fig:triangle}, 
resides in the interior of the symmetry triangle and 
connects the U(5) and SU(3) vertices. 

\section{PDS and QDS in a first-order quantum phase transition} 

Quantum phase transitions (QPTs) are qualitative changes in the 
properties of a physical system induced by a variation 
of parameters in the quantum Hamiltonian. 
Such structural changes are currently of great interest in different 
branches of physics~\cite{carr10}. 
The competing interactions in the Hamiltonian that drive these 
ground-state phase transitions can affect dramatically the nature of 
the dynamics and, in some cases, lead to the emergence of quantum 
chaos~\cite{Emar03,MacLev11,LevMac12,MacLev14}.
Here we show that PDS and QDS can characterize the remaining 
regularity in a system undergoing a QPT, 
amidst a complicated environment of other states~\cite{MacLev14}. 

Focusing on the dynamics at the critical-point of a first-order QPT 
between spherical and deformed shapes, 
the relevant IBM Hamiltonian~\cite{Lev06}, upto a scale, 
can be taken to be the second term of Eq.~(\ref{HPSsu3}), 
$\hat{H}_{\rm cri}=P^{\dagger}_{2}\cdot \tilde{P}_{2}$. 
The latter has 
the SU(3) basis states of Eq.~(\ref{ggamband}) and 
the following U(5) basis states
\bsub
\ba
\label{ePDSu5L0}
\vert N, n_d=\tau=L=0\rangle \;\;\;\; 
&&E = 0 ~,\\[2mm]
\vert N, n_d=\tau=L=3\rangle \;\;\;\; 
&&E = 6( 2N -1)~,
\label{ePDSu5L3}
\ea
\label{ePDSu5}
\esub
as solvable eigenstates, 
while all other states are mixed with respect to both U(5) and SU(3).
As such, $\hat{H}_{\rm cri}$ exhibits a coexistence of SU(3)-PDS and 
U(5)-PDS~\cite{Lev07}.

The classical limit of the Hamiltonian 
is obtained through the use of Glauber 
coherent states. This amounts to 
replacing $(s^{\dagger},\,d^{\dagger}_{\mu})$ by 
c-numbers $(\alpha_{s}^{*},\,\alpha_{\mu}^{*})$ rescaled 
by $\sqrt{N}$ and taking $N\rightarrow\infty$, with $1/N$ playing the 
role of $\hbar$~\cite{Hatch82}. 
Setting all momenta to zero, yields the classical 
potential $V(x,y)$, 
which coincides with the surface of Eq.~(\ref{enesurf}), 
with $(\beta,\gamma)$ as polar coordinates. 
The classical dynamics constraint to $L=0$, 
can be depicted conveniently via 
Poincar\'e surfaces of sections in the plane $y=0$, 
plotting the values of $(p_x,x)$ 
each time a trajectory intersects the plane. 
Regular trajectories are bound to toroidal manifolds within the 
phase space and their intersections with the plane of section lie 
on 1D curves (ovals). In~contrast, chaotic trajectories randomly 
cover kinematically accessible areas of the section.

The Poincar\'e sections associated with the classical 
critical-point Hamiltonian 
are shown in Fig.~10 
for representative energies. 
The bottom panel displays the classical potential which has two degenerate 
spherical and deformed minima. 
The dynamics in the region of the deformed minimum 
is robustly regular. The trajectories form a single island 
and remain regular even 
at energies far exceeding the barrier height $V_{\rm bar}$. 
In contrast, the dynamics in the region of the spherical minimum shows 
a change with energy 
from regularity to chaos, until complete chaoticity is reached 
near the barrier top. 
The clear separation between regular and chaotic dynamics, 
associated with the two minima, persists all the way to the 
barrier energy, $E=V_{\rm bar}$, where the two regions just touch. 
At $E > V_{\rm bar}$, a layer of chaos develops 
in the deformed region and gradually dominates the surviving 
regular island for $E\gg V_b$.
\begin{figure}[!t]
\begin{minipage}{10pc}
\begin{center}
\includegraphics[width=3.5cm]{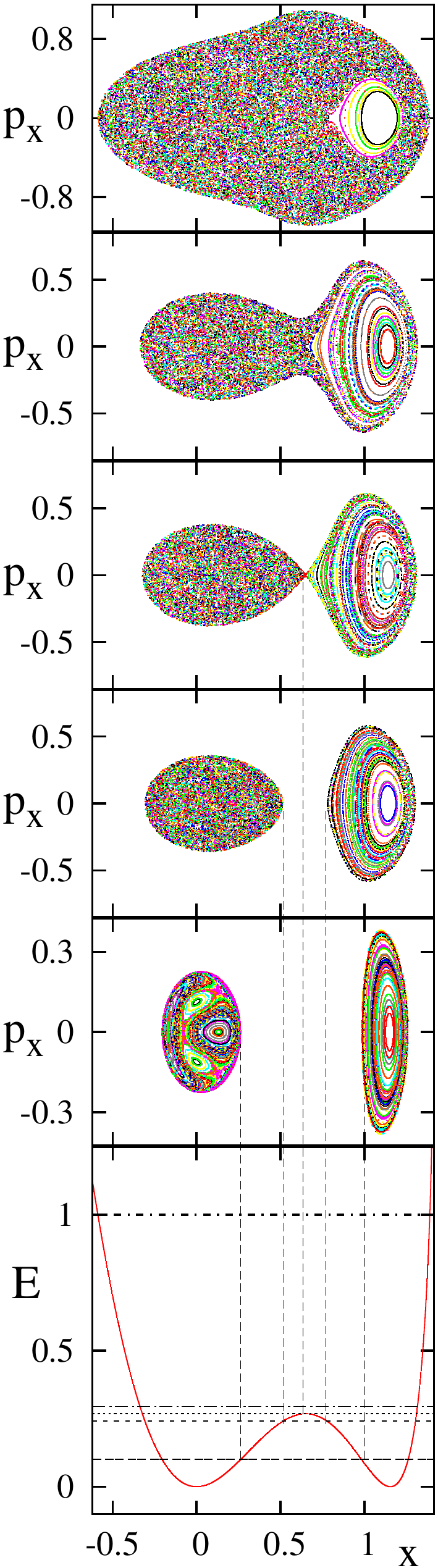}
\end{center}
\end{minipage}
\hspace{1cm}
\begin{minipage}{23pc}
\vspace{-0.5cm}
\caption{
\small
Poincar\'e sections at the critical point of the first-order QPT 
(upper five rows). The bottom row displays the classical potential. 
The five energies at which the sections were calculated consecutively, 
are indicated by horizontal lines. Adapted from~\cite{MacLev14}.}
\vspace{0.8cm}
\begin{center}
\includegraphics[height=7cm]{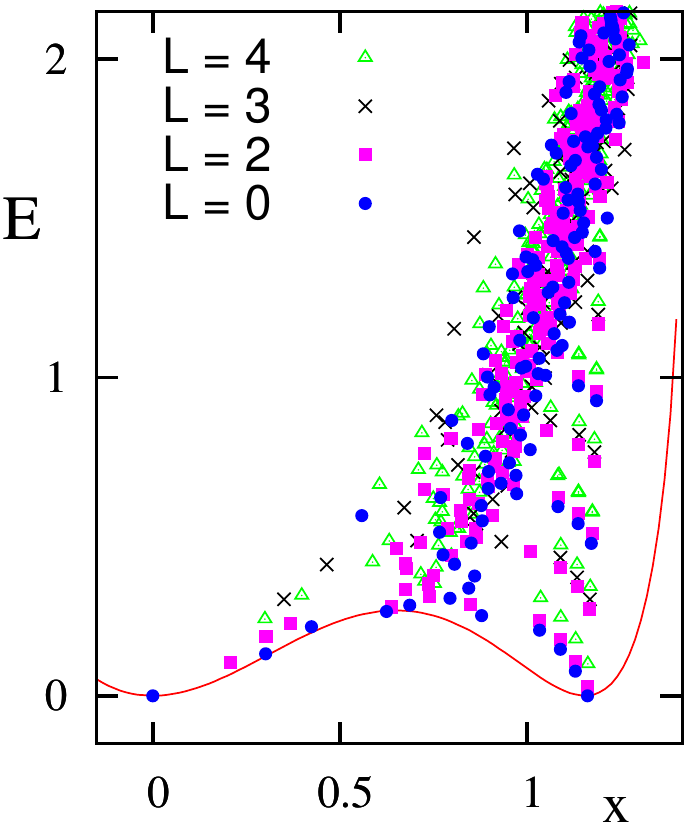}
\caption{
\small
Peres lattices $\{x_i,E_i\}$ for $L=0,2,3,4$ and $N=50$ eigenstates
of the critical-point Hamiltonian, discussed in the text.
The lattices are overlayed on the corresponding
classical potential. Adapted from~\cite{MacLev14}.}
\end{center}
\end{minipage}
\end{figure}
\begin{figure}[!t]
\begin{center}
\includegraphics[width=0.49\linewidth]{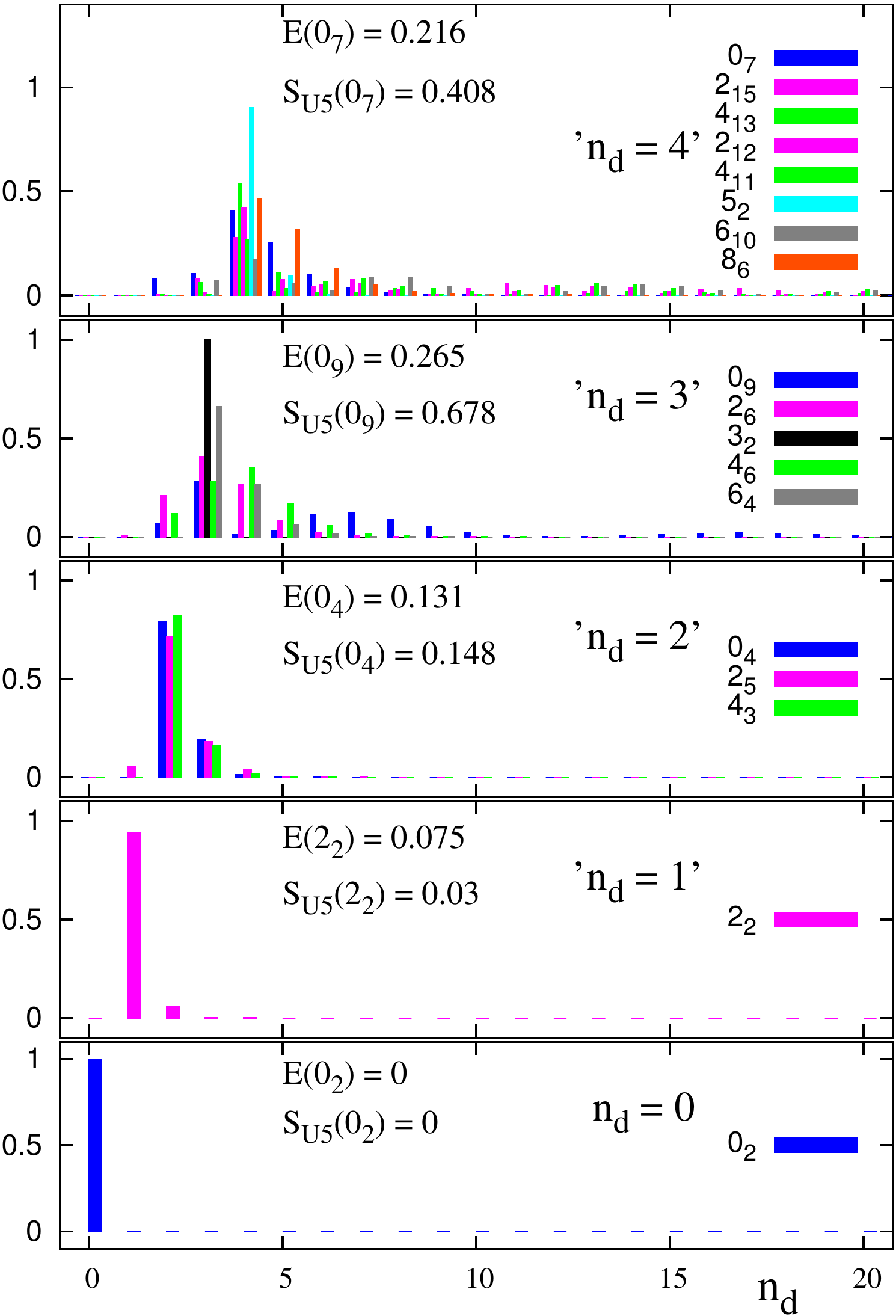}
\includegraphics[width=0.49\linewidth]{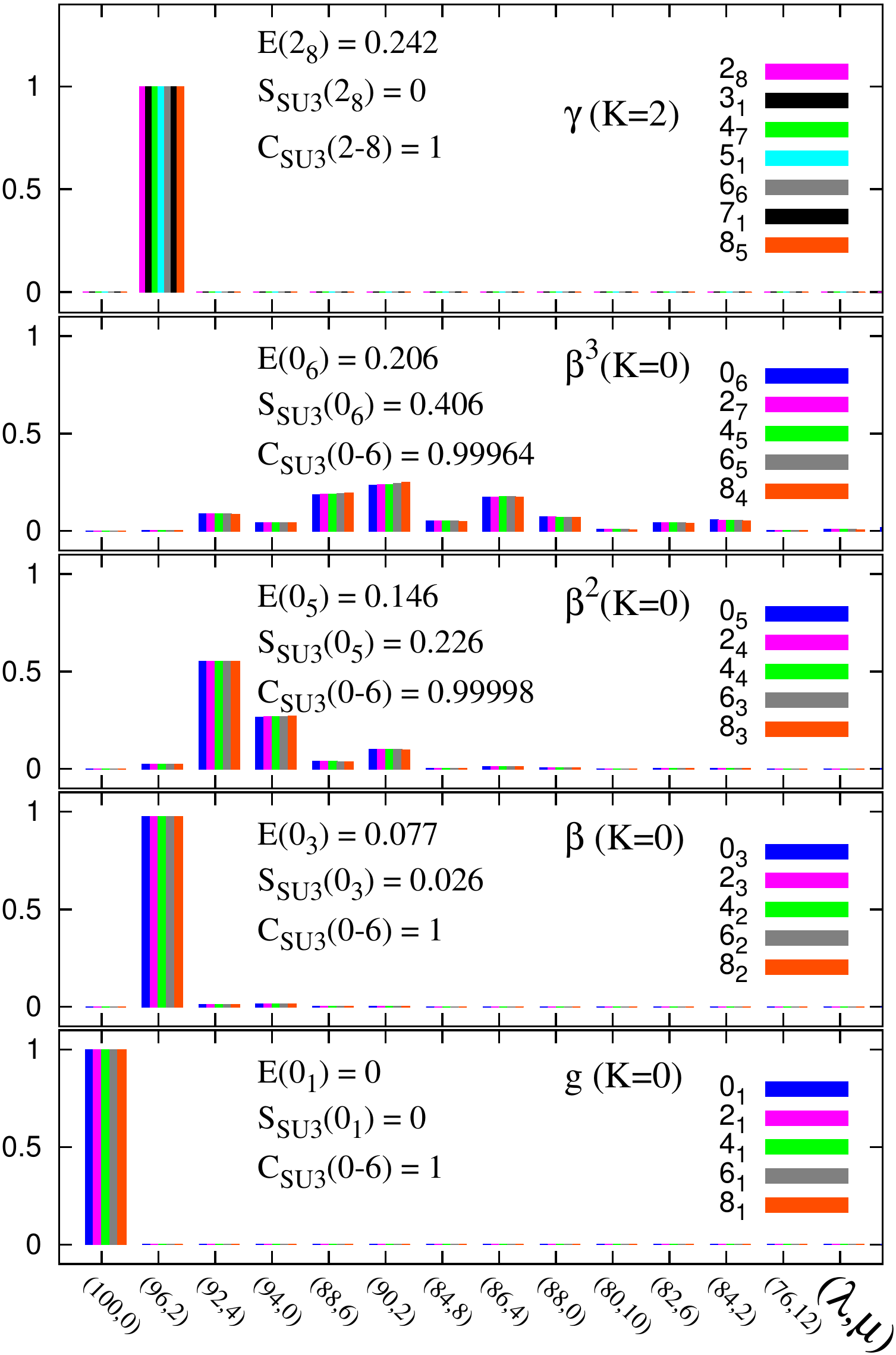}
\end{center}
\caption{
\small
U(5) $n_d$-probability distribution, $P_{n_d}^{(L_i)}$~(\ref{Pnd}) 
[left column], and SU(3) $(\lambda,\mu)$-probability distribution, 
$P_{(\lambda,\mu)}^{(L_i)}$~(\ref{Plammu}) [right column], for selected 
eigenstates of the critical-point Hamiltonian, with $N=50$. 
The U(5) Shannon entropy, $S_{\rm U5}(L_i)$~(\ref{Pnd}),
and SU(3) correlator, $C_{\rm SU3}(0{\rm -}6)$ 
are indicated for spherical and deformed type of states, respectively. 
Adapted from~\cite{MacLev14}.}
\label{fig6}
\end{figure}

The quantum manifestations of such an 
inhomogeneous phase space structure, 
can be studied by 
the method of Peres lattices~\cite{Peres84}. 
The latter are constructed by plotting the expectation 
values $O_i = \bra{i}\hat{O}\ket{i}$ of an arbitrary operator, 
$[\hat{O},\hat{H}]\neq 0$, versus the energy $E_i = \bra{i}\hat{H}\ket{i}$ 
of the Hamiltonian eigenstates $\ket{i}$. 
The lattices $\{O_i,E_i\}$ corresponding to 
regular dynamics display an ordered pattern, while chaotic 
dynamics leads to disordered meshes of 
points~\cite{Peres84,Stran09}. 
In the present analysis, we choose the Peres operator to be 
$\hat{n}_d$, whose expectation value 
is related to the coordinate $x$ in the classical potential.
The lattices 
$\{x_i,E_i\}$, with $x_i \equiv \sqrt{2 \bra{i}\hat{n}_d\ket{i}/N}$, 
can then distinguish regular from irregular states 
and associate them with a given region in phase~space.

The Peres lattices  
corresponding to 
($N\!=\!50,\,L\!=\!0,2,3,4$) eigenstates of $\hat{H}_\mathrm{cri}$
are shown in Fig.~11, overlayed on the classical  potential.
They disclose 
regular sequences of states localized within and above the deformed well. 
They are comprised of rotational states with $L=0,2,4,\ldots$ 
forming regular $K\!=\!0$ bands and 
sequences $L=2,3,4,\ldots$ 
forming $K=2$ bands. 
Additional $K$-bands (not shown in Fig.~11), 
corresponding to multiple $\beta$ and $\gamma$ vibrations 
about the deformed shape, can also be identified.
The states in each regular band 
share a common structure, to be discussed below.
Such ordered band-structures persist to energies above the barrier and 
are not present in the disordered (chaotic) portions 
of the Peres lattice. 
At low-energy, in the vicinity of the spherical well, one can also detect 
multiplets of states with $L=0$, $L=2$ and $L=0,2,4$, typical of 
quadrupole excitations of a spherical shape.

An important clue on the nature of the surviving regular sequences
of selected states, in the presence of more complicated type of eigenstates, 
comes from a symmetry analysis of their wave functions. 
The left column of Fig.~12 shows the U(5) $n_d$-probabilities, 
$P_{n_d}^{(L_i)}$~(\ref{Pnd}), for eigenstates of $\hat{H}_{\rm cri}$, 
selected on the  basis of having 
the largest components with $n_d=0,1,2,3,4$,
within the given $L$ spectra. 
The states are arranged into panels labeled 
by `$n_d$' to conform with the structure of the $n_d$-multiplets of the 
U(5) DS limit. The normalized 
U(5) Shannon entropy $S_{\rm U5}(L_i)$, Eq.~(\ref{Pnd}), 
is indicated for representative eigenstates. 
In particular, the zero-energy $L\!=\!0^{+}_2$ state 
is seen to be a pure $n_d\!=\!0$ state, 
with $S_{\rm U5}\!=\!0$,  
which is the solvable U(5)-PDS eigenstate of Eq.~(\ref{ePDSu5L0}). 
The state $2^{+}_2$ has a pronounced $n_d\!=\!1$ 
component~(96\%) and the states ($L=0^{+}_4,\,2^{+}_5,\,4^{+}_3$) 
in the third panel, have a pronounced $n_d\!=\!2$ component and
a low value of $S_{\rm U5}< 0.15$.
All the above states with $`n_d\leq 2'$ have a dominant single $n_d$ 
component, and hence qualify as `spherical' type of states.
These are the lowest left-most states in the 
Peres lattices of Fig.~11, mentioned above.
In contrast, the states in the panels `$n_d=3$' and `$n_d=4$' of Fig.~12, 
are significantly fragmented. A notable exception is 
the $L=3^{+}_2$ state, which is the solvable U(5)-PDS state of 
Eq.~(\ref{ePDSu5L3}) with $n_d=3$.
The existence in the spectrum of specific spherical-type of states with 
either $P_{n_d}^{(L)}\!=\!1$ $[S_{\rm U5}(L)\!=\!0]$ 
or $P_{n_d}^{(L)}\approx 1$ $[S_{\rm U5}(L)\approx 0]$, exemplifies 
the presence of an exact or approximate U(5) PDS at the critical-point.

The states shown on the right column of Fig.~12 
have a different character. They 
belong to the five lowest regular sequences 
seen in the Peres lattices of Fig.~11, in the region $x\geq 1$. 
They have a broad $n_d$-distribution, hence are qualified as
`deformed'-type of states, forming rotational bands:
$g(K\!=\!0),\,\beta(K\!=\!0),\,\beta^2(K\!=\!0),
\,\beta^3(K\!=\!0)$ and $\gamma(K\!=\!2)$.
Each panel depicts the SU(3) $(\lambda,\mu)$-distribution, 
$P_{(\lambda,\mu)}^{(L_i)}$ 
for the band members, 
the normalized SU(3) Shannon entropy $S_{\rm SU3}(L)$~(\ref{Plammu}) 
for the bandhead state, and the Pearson correlator 
$C_{\rm SU3}(0_i{\rm -}6)$ defined in Section~6.
The ground $g(K\!=\!0)$ and $\gamma(K\!=\!2)$  
bands are pure $[S_{\rm SU3}=0$] 
with $(\lambda,\mu) = (2N,0)$ and $(2N-4,2)$ SU3) character, respectively. 
These are the solvable bands of 
Eq.~(\ref{ggamband}) with SU(3) PDS. 
The non-solvable $K$-bands are mixed with respect 
to SU(3) in a coherent, $L$-independent, manner, hence 
exemplify SU(3)-QDS. 
As expected, we find $C_{\rm SU3}(0_i{\rm -}6)\approx 1$ for 
these $K$-bands. The persisting regular U(5)-like [SU(3)-like] multiplets 
reflect the geometry of the classical Landau potential, as they are associated
with the different spherical (deformed) minimum.
One can use the corresponding measures of PDS and QDS as fingerprints of 
the QPT, not only at the critical point, but also throughout the coexistence 
region, where the two minima interchange~\cite{MacLev14}.

\section{Concluding remarks}

The many examples of PDS and QDS, discussed in the present 
contribution, demonstrate that these intermediate-symmetries are more 
abundant than previously recognized. Contrary to naive expectations, 
the symmetry triangle appears to encompass important elements of 
symmetry and ``not all is lost'' inside it.
Although, the examples considered were presented in the framework of 
a bosonic model, it is important to emphasize that these symmetry 
concepts are applicable to any many-body problem (bosons and fermions) 
endowed with an algebraic structure. Examples of many-body Hamiltonians 
with fermionic PDS and QDS are known~\cite{escher00,RosenRowe03,isa08}. 
The PDS algorithms discussed, for constructing Hamiltonians with PDS, 
are applicable to any semi-simple algebra and can be extended to coupled 
algebraic structure, $G_1\times G_2$~\cite{talmi97,levgin00}. 
Attempts are under way to extend the PDS notion to Bose-Fermi symmetries 
and supersymmetries~\cite{ijl14}.

An important virtue of the PDS algorithms is their ability to 
incorporate and provide a selection criterion for higher-order 
terms~\cite{lev13}. 
Such terms are needed for an accurate 
description of the data and for extensions of ab-initio and 
beyond-mean-field methods to larger systems, which necessitate 
a strategy to deal with A-body effective interactions 
and proliferation of parameters.
On one hand, the PDS approach 
allows more flexibility by relaxing the constrains of an exact dynamical 
symmetry (DS). On the other hand, the PDS approach picks 
particular symmetry-breaking terms which do not destroy results previously 
obtained with a DS for a segment the spectrum. The PDS construction is 
implemented order by order, yet the scheme is non-perturbative in the sense 
that the non-solvable states can experience strong symmetry-breaking.
These virtues generate an efficient tool which can greatly enhance 
the scope of algebraic modeling of dynamical systems.

Correlated quantum many-body systems often display an astonishing 
regular excitation patterns which raises a fundamental question, namely, 
how simplicity emerges out of complexity in such circumstances. 
The simple patterns show up amidst 
a complicated environment of other states. 
It is natural to associate the ``simple'' states with a symmetry 
that protects their purity and special character. This symmetry, 
however, is shared by only a subset of states, and is broken in the remaining 
eigenstates of the same Hamiltonian. It thus appears that realistic 
quantum many-body Hamiltonians can accommodate simultaneously 
eigenstates with different symmetry character. 
The symmetry in question cannot be exact but only partial or ``apparent''. 
These are precisely the defining ingredients of PDS and QDS. 
These novel concepts of symmetries 
can thus offer a possible clue in addressing the ``simplicity out 
of complexity'' challenge.

\ack
I thank P. Van Isacker, J.E. Garc\'ia Ramos, C. Kremer, N. Pietralla 
and M. Macek for a fruitful collaboration on the topics discussed and 
F. Iachello, P. Cejnar and R.F. Casten 
for insightful discussions. 
This work is supported by the Israel Science Foundation.


\smallskip

\end{document}